\documentclass[12pt]{article}

\usepackage{url}
\usepackage{bbm}
\usepackage{mathtools}
\usepackage{amssymb}
\usepackage{amsthm}
\usepackage{empheq}
\usepackage{latexsym}
\usepackage{enumitem}
\usepackage{eurosym}
\usepackage{dsfont}
\usepackage{appendix}
\usepackage{color} 
\usepackage{hyperref}
\usepackage[utf8]{inputenc}
\usepackage[T1]{fontenc}
\usepackage{geometry}
\usepackage{multirow}
\usepackage{lmodern}
\usepackage{anyfontsize}
\usepackage{stmaryrd}
\usepackage{float}
\usepackage{bookmark}
\usepackage{subcaption}

\usepackage{graphicx}
\usepackage{epstopdf}

\allowdisplaybreaks

\def \O{\mathbb{O}}
\def \R{\mathbb{R}}
\definecolor{red}{rgb}{0.7,0.15,0.15}
\definecolor{green}{rgb}{0,0.5,0}
\definecolor{blue}{rgb}{0,0,0.7}
\hypersetup{colorlinks, linkcolor={red},citecolor={green}, urlcolor={blue}}
            
\makeatletter \@addtoreset{equation}{section}

\newtheorem{theorem2}{Theorem}[section]

\newtheorem{remark}[theorem2]{Remark}

\DeclareUnicodeCharacter{014D}{\=o}
\allowdisplaybreaks

\title{An approximate solution for options market-making in high dimension\footnote{This work benefits from the financial support of the Chaires Analytics and Models for Regulation, Financial Risk and Finance and Sustainable Development. Bastien Baldacci and Joffrey Derchu gratefully acknowledge the financial support of the ERC Grant 679836 Staqamof. The authors would like to thank Mathieu Rosenbaum (Ecole Polytechnique), Dylan Possamai (ETH Zurich), Greg Sidier (G-Research) and Olivier Guéant (Université Paris-1 Panthéon-Sorbonne) for numerous fruitful discussions. In particular, Mathieu Rosenbaum deserves warm thanks for his careful reading of the paper and his many suggestions to improve its quality.}}

\author{Bastien {\sc Baldacci}\footnote{\'Ecole Polytechnique, CMAP, 91128, Palaiseau, France,  bastien.baldacci@polytechnique.edu.} \and Joffrey {\sc Derchu}\footnote{\'Ecole Polytechnique, CMAP, 91128, Palaiseau, France,  joffrey.derchu@polytechnique.edu.} \and Iuliia {\sc Manziuk}\footnote{\'Ecole Polytechnique, CMAP, 91128, Palaiseau, France,  iuliia.manziuk@polytechnique.edu.}}

\begin{document}

\maketitle
\begin{abstract}
Managing a book of options on several underlying involves controlling positions of several thousands of financial assets. It is one of the most challenging financial problems involving both pricing and microstructural modeling. An options market maker has to manage both long- and short-dated options having very different dynamics. In particular, short-dated options inventories cannot be managed as a part of an aggregated inventory, which prevents the use of dimensionality reduction techniques such as a factorial approach or first-order Greeks approximation. In this paper, we show that a simple analytical approximation of the solution of the market maker's problem provides significantly higher flexibility than the existing algorithms designing options market making strategies. \\

\noindent{\bf Keywords:} Option market making, stochastic control, partial differential equations. 

\end{abstract}

\section{Introduction}\label{Section Introduction}

After the electronification of delta-one trading, where high-frequency trading companies provide the vast majority of the liquidity on several thousands of assets, systematic options trading seems to be the next main challenge in quantitative trading. For assets listed in a central limit order book, as in the equity world, execution and market making are carried out using algorithms. However, for less mature markets such as a great proportion of fixed income securities, systematic market making activities are driven by request-for-quote (RFQ for short) systems: the client sends a request to obtain a buy or sell price, for a given quantity of a security, to one or several market makers, who propose prices based on their current positions. Given the prices, the client accepts or refuses one or several transactions. On OTC markets, such as the corporate bonds market, the proportion of the volume traded with electronic market makers is increasing. \\

For more than three decades, the optimal market making problem on cash markets has been the object of many academic studies. The two primary references are \cite{grossman1988liquidity, ho1981optimal}. In \cite{grossman1988liquidity}, the authors proposed a simple three-period economic model representing the interaction between market makers and market-takers and analyzed the equilibrium state. In \cite{ho1981optimal}, the authors studied the behavior of a market maker facing a stochastic demand and an inventory risk and obtained his optimal strategy using the stochastic optimal control theory. In the well-known paper of Avellaneda and Stoikov \cite{avellaneda2008high} inspired by this framework, they proposed a model applicable for market making on the order-driven market at the high-frequency. However, due to the continuous nature of the market maker's spreads, and the assumption that the underlying asset is a diffusion process, this model is more suited to quote-driven markets such as corporate bonds market. \\

By providing a rigorous analysis of the stochastic control problem of \cite{avellaneda2008high}, the authors of \cite{gueant2013dealing} show, in the case of a CARA utility function, that the market maker's problem boils down to a system of linear ordinary differential equations. A large part of the contribution to the market making literature comes from works of Cartea and Jaimungal, who enriched the initial model by introducing alpha signals, ambiguity aversion, competition with other agents, see, for example, \cite{cartea2017algorithmic, cartea2018enhancing, cartea2016incorporating, cartea2015algorithmic}. In these works, they consider a risk-adjusted expectation maximization. As shown in \cite{manziuk2019optimal}, the solution of such formulation can also be obtained through CARA utility maximization after a suitable intensity function transformation. More recently, multi-asset market making, still on linear markets, has been addressed through reinforcement learning techniques, see \cite{gueant2019deep}, and dimensionality reduction techniques, as in \cite{bergault2019size}. \\

Regardless of how rich is the part of academic literature considering linear markets, the part studying optimal market making on options is far less extensive. A reasonable market making model for options has to take into account a lot of stylized facts. First, option market makers trade simultaneously derivatives and the corresponding underlying, which implies the construction of more complex trading strategies taking into account, for example, the Delta-Vega hedging. Consequently, one needs to impose a factorial stochastic volatility model, possibly with jumps, on the underlying asset. Second, option market makers need to manage several thousands of positions, which lead to very high-dimensional problems that cannot be solved using classical numerical schemes. Even if machine learning techniques are used, involving, for example, deep reinforcement learning methods (see \cite{gueant2019deep, weinan2017deep}), the computation time can still be an obstacle. The market maker has to answer a request from a client in a given time, which can be insufficient to recalibrate the model if some parameter changes need to be applied (for example, the correlation structure). Finally, when dealing with short maturity options, the market maker has to manage the positions individually to avoid sudden high exposure due to the Gamma of a specific position. This specificity prevents the use of some dimensionality reduction techniques. \\

In the existing academic literature, options market making is addressed in \cite{baldacci2019algorithmic, el2015stochastic, stoikov2009option}. In \cite{stoikov2009option}, the authors consider three different settings for a market maker managing a single option and its underlying. The first setting is a complete market with continuous trading in the perfectly liquid underlying. The second is a complete market with an illiquid underlying where the market maker sets bid and ask quotes in the option and the stock. The third is an incomplete market with residual risks due to stochastic volatility and overnight jumps in the stock price. In \cite{el2015stochastic}, the authors consider a market maker in charge of a single option in a framework à la Avellaneda-Stoikov, where an underlying follows a one-factor stochastic volatility model, and the market maker is always Delta-hedged. They provide optimal bid and ask quotes for the option taking into account the risk of model misspecification. Finally, in \cite{baldacci2019algorithmic}, the authors consider a perfectly Delta-hedged market maker in charge of a book of options with long maturities, whose prices are driven by a stochastic volatility model. The only risk factor comes from the Brownian motion driving the volatility of the underlying. Using a first-order approximation of the Vega of the portfolio, they show that the problem of an options market maker boils down to a three-dimensional Hamilton-Jacobi-Bellman (HJB) equation, which can be solved using classical finite difference schemes. By linearizing the value function of the market maker around the Vegas at the initial time, they provide a way to relax the constant Vega assumption. However, the disadvantage of this approach is its time-consumption due to the necessity to simulate inventory trajectories. Moreover, the constant Vega assumption, making the control problem time-inconsistent, is only valid for a market maker in charge of long-dated options where possible jumps in the underlying do not influence the global risk position drastically. Finally, if one adds other Greeks such as Vanna and Vomma, the model becomes hardly tractable as the HJB equation is in dimension $5$.   \\

In this article, our goal is to propose a market making algorithm that considers the three specificities mentioned above, more flexible and applicable in practice. To this end, we consider a market maker in charge of a book of options on different underlyings. The assets follow a one-factor stochastic volatility model with jumps, and the Brownian motions driving the underlying and the volatility of each asset are correlated. We first consider the case of a perfectly Delta-hedged market maker who manages his volatility Greeks, namely the Vega, the Vanna, and the Vomma, for all his positions. Inspired by \cite{evangelista2018new}, we approximate the jump-diffusion HJB equation corresponding to the optimization problem of the market maker with an elliptic Partial Differential Equation (PDE for short). Using an ansatz quadratic in the inventories, we approximate the value function by a system of non-linear PDEs, which can be easily solved via classical numerical methods for a small number of assets. For a number of underlyings above two, we recast the ansatz by adding a non-local term, enabling the use of the Deep Galerkin method as in \cite{hirsa2020unsupervised} to solve the system of PDEs rapidly due to its simple non-linearity.\\

The method presented in this paper has several advantages. First, contrary to \cite{el2015stochastic} and similarly to~\cite{baldacci2019algorithmic}, the market maker can design trading strategies on a high number of options. Contrary to \cite{baldacci2019algorithmic}, the market maker controls each position individually, which is particularly important for short-dated options that must be managed one by one. Moreover, it enables us to reproduce classic option market making behavior where one option is hedged with another. Second, we allow continuous updates of the Greeks (Delta, Vega, Vanna, Vomma) of each option, and the dependence of the intensities of orders arrival on the dynamics of the underlying and its stochastic volatility. This is a major improvement compared to \cite{baldacci2019algorithmic}, as the quotes of the market maker are adjusted dynamically with respect to the evolution of both an underlying and stochastic volatility, allowing the problem to be solved in a time-consistent way. Third, we can use a model for the underlying dynamics with an arbitrary number of factors without increasing the computation time. We show numerically how this algorithm outperforms the one in \cite{baldacci2019algorithmic} in terms of average PnL for a portfolio of options, where Vegas vary significantly. \\

The paper has the following structure: in Section \ref{sec_framework}, we present the framework of options market making and the corresponding optimization problem faced by the market maker. In Section \ref{Ansatz}, we show how to simplify the problem by approximating the value function. Finally, Section \ref{sec_numerics} is devoted to numerical experiments. 

\section{Framework}\label{sec_framework}

\subsection{The option book}

We consider a filtered probability space $(\Omega,\mathcal{F},\mathbb{P})$ where all stochastic processes are defined, and a time horizon $T>0$. We consider $d>1$ stocks with the following one factor stochastic volatility dynamics with jumps: 
\begin{align}\label{eq_stock_dynamics}
\left\{
    \begin{array}{ll}
        dS^i_{t} = b_{\mathbb{P}}^i (t,S_t^i) dt +\sigma^i(t,S_t^i, \nu_t^i)dW_{t}^{i,S} + \int_{\mathbb{R}} Z^i(dt,dz)  \\
        d\nu^i_{t}=a^i_{\mathbb{P}}(t,\nu^i_t)dt+v^i_{\mathbb{P}}(t,\nu_t^i)dW_{t}^{i,\nu},
    \end{array}
\right.
\end{align}
where $(W_{t}^{i,S},W_{t}^{i,\nu})_{t\in \mathbb{R}^{+}}$ is a couple of Brownian motions with quadratic covariation given by the coefficients $\rho^{i}=\frac{d\langle W^{i,S},W^{i,\nu}\rangle}{dt}  \in (-1,1)$, and $a^i_{\mathbb{P}},b^i_{\mathbb{P}},v^i_{\mathbb{P}},\sigma^i$ are such that the SDEs \eqref{eq_stock_dynamics} admit a unique strong solution\footnote{In particular, for the sake of readability, we assume that there is no correlation between the volatility process of an asset and the variations of another asset. This assumption can be directly relaxed.}. The processes $Z^i(dt,dz)$ are marked point processes independent from the Brownian motions, with intensity kernels $\kappa_t^i(dz)$. We also assume that there exists covariance matrices $\Sigma^S,\Sigma^{\nu} \in \mathcal{M}_d(\mathbb{R})$ which correspond to the correlation structure of the stocks and the stochastic volatility in the option book. There also exists a risk-neutral probability measure $\mathbb{Q}$ such that 
\begin{align*}
\left\{
    \begin{array}{ll}
        dS^i_{t} = \sigma^i(t,S_t^i, \nu_t^i)d\hat{W}_{t}^{i,S} + \int_{\mathbb{R}}Z^i(dt,dz)  \\
        d\nu^i_{t}=a^i_{\mathbb{Q}}(t,\nu^i_t)dt+v^i_{\mathbb{Q}}(t,\nu_t^i)d\hat{W}_{t}^{i,\nu},
    \end{array}
\right.
\end{align*}
where $(\hat{W}_{t}^{i,S},\hat{W}_{t}^{i,\nu}), i\in \{1,\dots,d\}$ are $\mathbb{Q}-$Brownian motions. 

\begin{remark}
As the reader will see in the following, by applying the ansatz detailed in Section \ref{Ansatz}, one can use a multi-factor stochastic volatility model for the underlying without increasing the complexity of the algorithm. For example, one can work with the well-known two-factor Bergomi model easily, see \cite{bergomi2008smile, bergomi2015stochastic}. 
\end{remark}

On every underlying $i\in \{1,\dots,d\}$ we consider a set of $N^i$ European options $\mathcal{O}^{i,j}$ of maturity~$T^{i,j}$, for $j\in \{1,\dots,N^i\}$. In the above one-factor model, we know that for all $(i,j)\in \{1,\dots,d\}\times \{1,\dots,N^i\}$, and all $t\in [0,T^{i,j}]$ such that $T\!<\!\min_{i,j} T^{i,j}$,  $\mathcal{O}_{t}^{i,j}=O^{i,j}(t,S_{t}^i,\nu^i_{t})$ where $O^{i,j}$ is a solution on $[0,T^{i,j})\times \mathbb{R}_{+}^2$ of the following partial differential equation under the probability $\mathbb{Q}$:
\begin{align*} 
\begin{split} 
0 = \,\, & \partial_{t}O^{i,j}(t,S^i,\nu^i)+a^i_{\mathbb Q}(t,\nu^i)\partial_{\nu^i}O^{i,j}(t,S^i,\nu^i) +\frac{1}{2}\big(\sigma^i (t,S^i,\nu^i)\big)^{2}\partial^{2}_{S^iS^i}O^{i,j}(t,S^i,\nu^i)\nonumber\\
& +\rho^{i,i} \nu^i_{\mathbb Q}(t,\nu^i)\sigma^i(t,S^i,\nu^i)\partial^{2}_{\nu^i S^i}O^{i,j}(t,S^i,\nu^i)+\frac{1}{2}\big(v_{\mathbb{Q}}^i(t,\nu^i)\big)^2\partial^{2}_{\nu^i\nu^i}O^{i,j}(t,S^i,\nu^i) \\
& + \int_{\mathbb{R}} \Big(O^{i,j}(t,S^i+\gamma^i(t,z),\nu^i)-O^{i,j}(t,S^i,\nu^i)\Big)\kappa^i(dz).
\end{split}
\end{align*}
As the time horizon $T$ is small compared to the maturity of the options (which can be from one day up to several years), the terminal condition of the PDEs does not have to be specified. In Section~\ref{sec_numerics}, numerical experiments are addressed using European call options but any other option with a path-independent payoff can be considered. We now define the market maker's problem. 

\subsection{The market maker's problem on OTC markets}

We consider a market maker in charge of providing bid and ask quotes for the $\sum_{i\in \{1,\dots,d\}} N^i$ options over the period $[0,T]$ where $T<\min_{i,j} T^{i,j}$. The bid and ask prices on the option $j\in \{1,\dots,N^i\}$ of stock $i\in \{1,\dots,d\}$ are defined, for transaction size $z$, by
\begin{align*}
    P_t^{i,j,b} = \mathcal{O}^{i,j}_t - \delta_t^{i,j,b}(z), \quad  P_t^{i,j,a} = \mathcal{O}^{i,j}_t + \delta_t^{i,j,a}(z), 
\end{align*}
where $\big(\delta_t^{i,j,b}(\cdot),\delta_t^{i,j,a}(\cdot)\big)\in \mathcal{A}$, where $\mathcal{A}$ is the set of uniformly bounded $\mathcal{F}$-predictable processes. They represent the spread on the bid or ask side of the option $\mathcal{O}^{i,j}$. The number or transactions on these options are defined by marked point processes $N^{i,j,b}(dt,dz),N^{i,j,a}(dt,dz)$, with almost surely no simultaneous jumps, whose respective intensity processes are given by
\begin{align*}
    \Lambda_t^{i,j,b}(S,\nu,dz)=\lambda^{i,j,b}\big(S,\nu,\delta_t^{i,j,b}(z)\big)\mu^{i,j,b}(dz), \quad \Lambda_t^{i,j,a}(S,\nu,dz)=\lambda^{i,j,a}\big(S,\nu,\delta_t^{i,j,a}(z)\big)\mu^{i,j,a}(dz).
\end{align*}
The couples $(\mu^{i,j,b},\mu^{i,j,a})$ are probability measures on $\mathbb{R}_+^\star$ modeling the distribution of transaction sizes for the options. Note that, in our framework, the intensities are allowed to depend on both the underlying and the stochastic volatility of the assets. \\

The market maker manages his inventory process on each option, that is
\begin{align*}
    dq_t^{i,j} = \int_{\mathbb{R}_+^\star}z \big(N^{i,j,b}(dt,dz) - N^{i,j,a}(dt,dz)\big).
\end{align*}
For the sake of simplicity, we represent the vector of inventories as follows:
\begin{align*}
    q^{\mathbf{T}} = \big(q^{1,1},\dots,q^{1,N^1},\dots,q^{d,N^d} \big) \in \mathcal{M}_{\sum_{l=1}^{d} N^l,1}(\mathbb{R}). 
\end{align*}
Assuming perfect Delta-hedging\footnote{This assumption can be relaxed by assuming that the market maker acts on the stock market. This way, the mean-variance objective function will take into account the Delta of the portfolio.}, the $\Delta$ of the portfolio on the $i$-th asset, $i\in \{1,\dots,d\}$, is given by 
\begin{align*}
    \Delta_t^i = \sum_{j\in \{1,\dots,N^i\}}\partial_{S^i}O^{i,j}(t,S_t^i,\nu_t^i)q_t^{i,j},\quad  \Delta_t = \sum_{i\in \{1,\dots,d\}} \Delta_t^i.
\end{align*}
The cash process of the market maker at time $t$ is defined by
\begin{align*}
    dX_t &= \!\!\!\! \sum_{\substack{(i,j)\in \\ \{1,\dots,d\} \times \\ \{1,\dots,N^i\}}}\!\!\!\!\!\Big(\int_{\mathbb{R}_+^\star}\!\!\!\! z\big(\delta_t^{i,j,b}(z)N_t^{i,j,b}(dt,dz) +  \delta_t^{i,j,a}(z)N_t^{i,j,a}(dt,dz)\big)  - \mathcal{O}_t^{i,j}dq_t^{i,j}\Big) \!\!+\!\!\!\! \sum_{i\in \{1,\dots,d\}} \big(S_t^i d\Delta_t^i + d\langle\Delta^{i},S^i\rangle_t\big).
\end{align*}
We finally define the Mark-to-Market value of the portfolio of the market maker as
\begin{align*}
    V_t = X_t - \sum_{i\in \{1,\dots,d\}} \Delta_t^i S_t^i + \sum_{(i,j)\in\{1,\dots,d\}\times\{1,\dots,N^i\}}q_t^{i,j}\mathcal{O}_t^{i,j}.
\end{align*}
For all $(i,j)\in\{1,\dots,d\}\times\{1,\dots,N^i\}$, the Vega, the Vomma and the Vanna of the option $\mathcal{O}_t^{i,j}$ are defined as
\begin{align*}
    &\mathcal{V}_t^{i,j}= \partial_{\sqrt{\nu^i}}O^{i,j}(t,S_t^i,\nu_t^i) = 2\sqrt{\nu^i}\partial_{\nu^i}O^{i,j}(t,S_t^i,\nu_t^i), \\
    &(\mathcal{VO})_t^{i,j}=\partial_{\sqrt{\nu^i}\sqrt{\nu^i}}O^{i,j}(t,S_t^i,\nu_t^i) = 4\nu^i \partial^2_{\nu^i \nu^i}O^{i,j}(t,S_t^i,\nu_t^i), \\
    & (\mathcal{VA})_t^{i,j}= \partial_{S\sqrt{\nu^i}}O^{i,j}(t,S_t^i,\nu_t^i) = 2\sqrt{\nu^i}\partial_{S\nu^i}O^{i,j}(t,S_t^i,\nu_t^i).
\end{align*}
We also define the vectors $e^{i,j}\in \mathbb{R}^{\sum_{l=1}^{d} N^l}$ where $e_k^{i,j}=\mathbf{1}_{\{k=\sum_{l=1}^{i-1} N^l+j\}}$ and $(e^1,\dots,e^d)$ as the canonical basis of $\mathbb{R}^d$. If we denote by $\Gamma_t^i = \frac{v_{\mathbb{P}}^i(t,\nu_t^i)}{2\sqrt{\nu_t^i}}\sum_{j\in \{1,\dots,N^i\}} q_t^{i,j} \mathcal{V}_t^{i,j}$, we can write the market maker's problem as
\begin{align}\label{pb_MM}
    \sup_{\delta \in \mathcal{A}}\mathbb{E}\Big[V_T - \frac{\gamma}{2}\sum_{(i,k)\in \{1,\dots,d\}^2}\int_0^T \Gamma_t^i \Gamma_t^k \Sigma^{\nu,i,k}dt  \Big].
\end{align}

Here we penalize the portfolio's total Vega. Any other penalization could be used, as long as it is quadratic in $q$. For example, this includes more complicated penalties linked to another position to hedge, or some target for the Greeks. We define the Hamiltonians
\begin{align*}
    H^{i,j,a}(S,\nu,p) = \sup_{\delta }\lambda^{i,j,a}(S,\nu,\delta) \big(\delta-p\big), \quad H^{i,j,b}(S,\nu,p) = \sup_{\delta }\lambda^{i,j,b}(S,\nu,\delta) \big(\delta-p\big),
\end{align*}
and the following processes $\mathcal{G}(t,S,\nu) \in \mathbb{R}^{\sum_{l=1}^{d} N^l}$ such that  
\begin{align*}
    &\mathcal{G}_{j}(t,S,\nu) =  \mathcal{V}_t^{k_j,j-\left(\sum_{l=1}^{k_j-1} N^l\right)}\frac{a^{k_j}_{\mathbb{P}}(t,\nu^{k_j})-a^{k_j}_{\mathbb{Q}}(t,\nu^{k_j})}{2\sqrt{\nu^{k_j}}} \\
    & \qquad \qquad + \rho^{k_j}(\mathcal{VA})_t^{k_j,j-\left(\sum_{l=1}^{k_j-1} N^l\right)}\frac{v^{k_j}_{\mathbb{P}}(t,\nu^{k_j})-v^{k_j}_{\mathbb{Q}}(t,\nu^{k_j})}{2\sqrt{\nu^{k_j}}}\sigma^{k_j}(t,S^{k_j},\nu^{k_j})\\
    & \qquad \qquad + (\mathcal{VO})_t^{k_j,j-\left(\sum_{l=1}^{k_j-1} N^l\right)}\frac{\big(v^{k_j}_{\mathbb{P}}(t,\nu^{k_j})\big)^2-\big(v^{k_j}_{\mathbb{Q}}(t,\nu^{k_j})\big)^2}{4\nu^{k_j}},
\end{align*}
where $k_j=i$ if $j\in \{\sum_{l=1}^{i-1} N^l,\dots,\sum_{l=1}^{i} N^l\}$, for $i\in \{1,\dots,d\}$. We also define $\mathcal{R}(t,S,\nu)\! \in\! \mathcal{M}_{\sum_{l=1}^{d}\!\! N^l,d}(\mathbb{R)}$ such that
\begin{align*}
   &  \mathcal{R}_{j,i}(t,S,\nu) = \frac{v_{\mathbb{P}}^i(t,\nu^i)}{2\sqrt{\nu_t^i}}\mathcal{V}_t^{i,j-\left(\sum_{l=1}^{k_j-1} N^l\right)}, \quad \text{for } j\in \{\sum_{l=1}^{i-1} N^l,\dots,\sum_{l=1}^{i} N^l\},i\in \{1,\dots,d\},
\end{align*}
and $0$ otherwise. Finally, denote the diffusion part of the HJB equation as
\begin{align*}
    \mathcal{L}(t,S,\nu,q,u) = & \sum_{i\in \{1,\dots,d\}} b^i_{\mathbb P}(t,S^i) \partial_{S^i}u(t,S,\nu,q) + \sum_{i\in \{1,\dots,d\}} a^i_{\mathbb P}(t,\nu^i) \partial_{\nu^i}u(t,S,\nu,q)\\
    & + \frac{1}{2}\sum_{(i,k)\in \{1,\dots,d\}^2} \partial_{S^i S^k}u(t,S,\nu,q) \sigma^i(t,S^i,\nu^i)\sigma^j(t,S^k,\nu^j)\Sigma^{S,i,k} \\
    & + \sum_{i\in \{1,\dots,d\}}\int_{\mathbb{R}}\kappa^i(dz)\Big(u\big(t,S + e^i\gamma^i(t,z),\nu,q\big)-u(t,S,\nu,q)\Big)  \\
    & + \frac{1}{2}\sum_{(i,k)\in \{1,\dots,d\}^2} \partial_{\nu^i \nu^j}u(t,S,\nu,q) v^i_{\mathbb{P}}(t,\nu^i)v^k_{\mathbb{P}}(t,\nu^k)\Sigma^{\nu,i,k} \\
    & + \sum_{i\in \{1,\dots,d\}}\partial_{\nu^i S^i}u(t,S,\nu,q) \rho^{i} v_{\mathbb{P}}^i (t,\nu^i) \sigma^i(t,S^i,\nu^i).
\end{align*}
The HJB equation associated to \eqref{pb_MM} with compact notations is
\begin{align}\label{HJB_Compact}
\begin{split}
0 = & \,\, \partial_t u(t,S,\nu,q) + \mathcal{L}(t,S,\nu,q,u) + q^{\mathbf{T}}\mathcal{G}(t,S,\nu) - \frac{\gamma}{2} q^{\mathbf{T}} \mathcal{R}(t,S,\nu) \Sigma^{\nu} \mathcal{R}^{\mathbf{T}}(t,S,\nu)q \\
& + \sum_{\substack{(i,j)\in \\ \{1,\dots,d\} \times \\ \{1,\dots,N^i\}}} \int_{\mathbb{R}_+}z H^{i,j,b}\Big(S,\nu,\frac{u(t,S,\nu,q)-u(t,S,\nu,q+z e^{i,j})}{z}\Big)\mu^{i,j,b}(dz) \\
& + \sum_{\substack{(i,j)\in \\ \{1,\dots,d\} \times \\ \{1,\dots,N^i\}}} \int_{\mathbb{R}_+}z H^{i,j,a}\Big(S,\nu,\frac{u(t,S,\nu,q)-u(t,S,\nu,q-z e^{i,j})}{z}\Big)\mu^{i,j,a}(dz).     
\end{split}
\end{align}
with terminal condition $u(T,S,\nu,q)=0$. The proof of existence and uniqueness of a viscosity solution to \eqref{HJB_Compact} associated to the control problem \eqref{pb_MM} relies on classic arguments of second order viscosity solutions with jumps, see for example \cite{barles2008second,bastien2020bid,bergault2019size}.

\section{Solving the market maker's problem with a system of non-linear PDEs}\label{Ansatz}

Equation \eqref{HJB_Compact} is intractable with classical numerical methods when dealing with several options on several underlyings. Notably, the method proposed in \cite{baldacci2019algorithmic} to overcome the constant Vega assumption requires Monte-Carlo simulations of high-dimensional inventory trajectories, which is very time-consuming. In this section, inspired by \cite{evangelista2018new}, we propose an approximation of the value function of the market maker, quadratic with respect to the vector of inventories to reduce the dimensionality of the problem. \\

A Taylor expansion at $0$ on the third variable with respect to $\epsilon$ gives
\begin{align*}
    & H^{i,j,b}\Big(S,\nu,\frac{u(t,S,\nu,q)-u(t,S,\nu,q+\epsilon z e^{i,j})}{z}\Big) +  H^{i,j,a}\Big(S,\nu,\frac{u(t,S,\nu,q)-u(t,S,\nu,q-\epsilon z e^{i,j})}{z}\Big) \\
     & = H^{i,j,b}(S,\nu,0) + H^{i,j,a}(S,\nu,0) + \epsilon\big( H^{' i,j,a}(S,\nu,0) - H^{i,j,b}(S,\nu,0)\big) \partial_{q}u(t,S,\nu,q) \\
    & +\frac{\epsilon^2}{2} \Big(H^{''i,j,a}(S,\nu,0)\big(\partial_q u(t,S,\nu,q)\big)^2 - zH^{'i,j,a}(S,\nu,0)\partial_{qq} u(t,S,\nu,q) \Big)  \\
    & + \frac{\epsilon^2}{2} \Big(H^{''i,j,b}(S,\nu,0)\big(\partial_q u(t,S,\nu,q)\big)^2 - zH^{'i,j,b}(S,\nu,0)\partial_{qq} u(t,S,\nu,q) \Big) + o (\epsilon^3),
\end{align*}
and by taking $\epsilon=1$, Equation \eqref{HJB_Compact} becomes
\begin{align}\label{HJB_ansatz}
    0 = \,\, & \partial_t u(t,S,\nu,q) + \mathcal{L}(t,S,\nu,q,u) +q^{\mathbf{T}} \mathcal{G}(t,S,\nu)  - \frac{\gamma}{2} q^{\mathbf{T}} \mathcal{R}(t,S,\nu) \Sigma^{\nu} \mathcal{R}^{\mathbf{T}}(t,S,\nu)q  \nonumber\\
    & + \sum_{\substack{(i,j)\in \\ \{1,\dots,d\} \times \\ \{1,\dots,N^i\}}} \int_{\mathbb{R}_+} \Bigg( H^{i,j,b}(S,\nu,0)  - H^{i,j,b}(S,\nu,0) \partial_{q}u(t,S,\nu,q) \nonumber\\
    & + \frac{1}{2} \Big(H^{''i,j,b}(S,\nu,0)\big(\partial_q u(t,S,\nu,q)\big)^2 - zH^{'i,j,b}(S,\nu,0)\partial_{qq} u(t,S,\nu,q) \Big)\Bigg)
    \mu^{i,j,b}(dz) \\
     & + \sum_{\substack{(i,j)\in \\ \{1,\dots,d\} \times \\ \{1,\dots,N^i\}}} \int_{\mathbb{R}_+}\Bigg (H^{i,j,a}(S,\nu,0) +  H^{' i,j,a}(S,\nu,0) \partial_{q}u(t,S,\nu,q) \nonumber\\
    & +\frac{1}{2} \Big(H^{''i,j,a}(S,\nu,0)\big(\partial_q u(t,S,\nu,q)\big)^2 - zH^{'i,j,a}(S,\nu,0)\partial_{qq} u(t,S,\nu,q) \Big) \Bigg)\mu^{i,j,a}(dz).\nonumber
\end{align}
In the following we will show how a simple ansatz, quadratic with respect to the vector of inventories, leads to significant simplifications. For the sake of the simplicity of the notation, assume  that $H^{i,j,a}=H^{i,j,b}=H^{i,j}$ (extension to asymmetric intensities is straightforward). By setting
\begin{align*}
    u(t,S,\nu,q)= \theta^0(t,S,\nu) + q^{\mathbf{T}}\theta^1(t,S,\nu) - q^{\mathbf{T}}\theta^2(t,S,\nu)q,
\end{align*}
where $\theta^0 \in \mathbb{R}, \theta^1 \in \mathbb{R}^{\sum_{l=1}^{d} N^l}, \theta^2 \in \mathcal{M}_{\sum_{l=1}^{d} N^l}(\mathbb{R})$ are solutions of the following system of non-linear PDEs: 
\begin{align}\label{system_pde}
\begin{cases}
0 = & \partial_t \theta^0(t,S,\nu) + \overline{\mathcal{L}}(t,\theta^0,\nu,S)  + 2 \sum_{(i,j)\in \{1,\dots,d\}\times \{1,\dots,N^i\}} H^{i,j}(S,\nu,0) \\
 & + \int_{\mathbb R_+}\Big(  2 z H^{'i,j}(S,\nu,0)\theta^2_{j,j}(t,S,\nu)
+ H^{'' i,j}(S,\nu,0)\big(\theta_j^1(t,S,\nu)\big)^2\Big)\mu^{i,j}(dz) \\
0 = & \partial_t \theta^1(t,S,\nu) +\overline{\mathcal{L}}(t,\theta^1,\nu,S)  + \mathcal{G}(t,S,\nu)+ 4 \theta^2(t,S,\nu)\text{diag}\big(H^{''}(S,\nu,0)\big) \theta^1(t,S,\nu) \\
0  = & \partial_t \theta^2(t,S,\nu) + \overline{\mathcal{L}}(t,\theta^2 ,\nu,S) - \frac{\gamma}{2} \mathcal{R}(t,S,\nu) \Sigma^{\nu} \mathcal{R}^{\mathbf{T}}(t,S,\nu) \\
& {} + 4 \theta^2(t,S,\nu) \text{diag}\big(H^{''}(S,\nu,0)\big)\theta^2(t,S,\nu),    
\end{cases}
\end{align}
where 
\begin{align*}
    \overline{\mathcal{L}}(t,\theta,\nu,S) = & \sum_{i\in \{1,\dots,N\}} b^i_{\mathbb P}(t,S^i) \partial_{S^i}\theta(t,S,\nu) + \sum_{i\in \{1,\dots,N\}} a^i_{\mathbb P}(t,\nu^i) \partial_{\nu^i}\theta(t,S,\nu) \\
    & {}+ \!\frac{1}{2} \!\!\sum_{(i,k)\in \{1,\dots,d\}^2}\!\!\!\!\!\!\!\!\!\!\! \big(\! \partial_{S^i S^j}\theta(t,S,\nu) \sigma^i\!(t,S^i\!,\nu^i)\sigma^k\!(t,S^k\!\!,\nu^k)\Sigma^{S\!,i\!,k} \!\!+\! \partial_{\nu^i \nu^k}\theta(t,S\!,\nu) v^i_{\mathbb{P}}(t,\nu^i)v^k_{\mathbb{P}}(t,\nu^k)\Sigma^{\nu,i,k}\!\big) \\
    &+ \!\!\!\sum_{i\in \{1,\dots,d\}}\!\!\bigg(\partial_{\nu^i S^i}\theta(t\!,S\!,\nu) \rho^{i} v_{\mathbb{P}}^i (t,\nu^i) \sigma^i(t\!,S^i\!,\nu^i)
   \! + \!\!\int_{\mathbb{R}}\!\kappa^i(dz)\Big(\theta\big(t\!,S\! +\! e^i\gamma^i(t,z)\!,\nu\big)\!\!-\!\!\theta(t\!,S\!,\nu)\Big)\bigg) . 
\end{align*}
and $\theta^0(T,S,\nu)=0,\theta^1(T,S,\nu)=\mathbf{0}_{\sum_{l=1}^{d} N^l,1},\theta^2(T,S,\nu)=\mathbf{0}_{\sum_{l=1}^{d} N^l}$. In system \eqref{HJB_ansatz}, one can note that the PDE with respect to $\theta^2$ is independent from the two others, which reduces the overall complexity. It can easily be solved for a small number of underlyings and a large number of options using finite difference schemes. Note that a higher order expansion does not yield a polynomial solution. However, it is possible to truncate the high degree terms to obtain a polynomial solution. This does not lead to a significant change of the value function or the controls if the penalty term is at most quadratic.\\

We now show some numerical applications of the methodology.

\section{Numerical results}\label{sec_numerics}

\begin{figure}[H]
\begin{center}
    \includegraphics[width=0.6\textwidth]{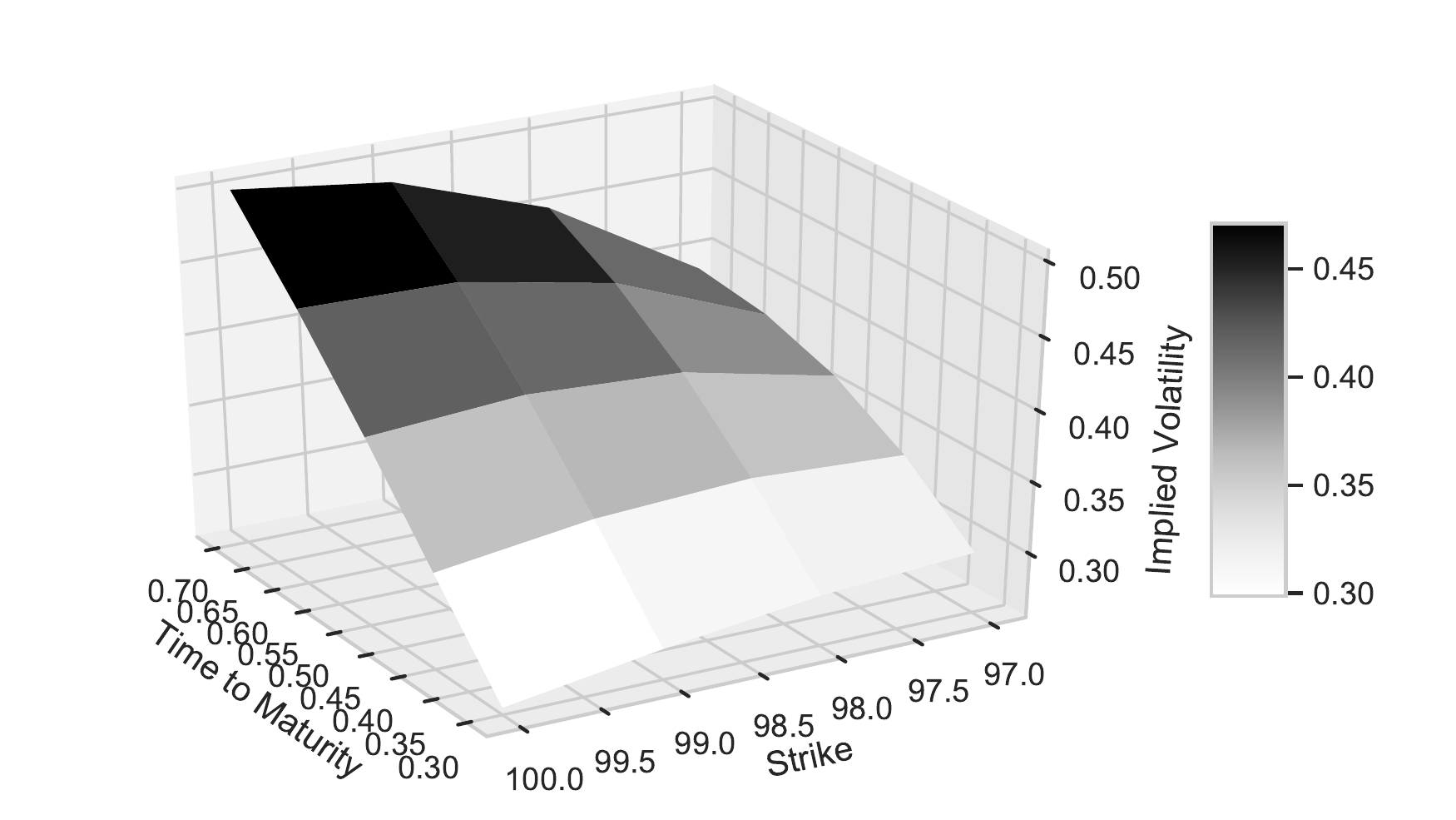}
    \caption{Implied volatility surface associated with the market parameters.}\label{impvolsurf}
\end{center}
\end{figure}

To perform a comparison with respect to the existing methods, we first recall the methodology of \cite{baldacci2019algorithmic}. In this article, the authors consider a market maker managing a book of options on a single underlying, and they suppose he is perfectly delta-hedged. We have the following set of market parameters:
\begin{itemize}
    \item $d=1,N^1=N=20$: there are $20$ call options on a single underlying.
    \item Stock price at time $t=0$: $S_0=100$\euro{}.
    \item Instantaneous variance at time $t=0$: $\nu_0=0.04\text{ year}^{-1}$.
    \item Heston model parameters: $b_{\mathbb{P}}(t,S)=\mu S$, $\sigma(t,S,\nu)=S\sqrt{\nu}$, $v_{\mathbb{P}}(t,\nu)=v_{\mathbb{Q}}(t,\nu)=\xi\sqrt{\nu}$, with $\xi=0.7\text{ year}^{-1}$.
    \item $a_{\mathbb{P}}(t,\nu)=\kappa_{\mathbb{P}}(\theta_{\mathbb{P}}-\nu),a_{\mathbb{Q}}(t,\nu)=\kappa_{\mathbb{Q}}(\theta_{\mathbb{Q}}-\nu)$, with $\kappa_{\mathbb{P}}=\kappa_{\mathbb{Q}}=2\text{ year}^{-1}$, $\theta_{\mathbb{P}}=\theta_{\mathbb{Q}}=0.04\text{ year}^{-1}$. 
    \item $Z(dt,dz)= 0$:  there is no jump in the dynamics of the underlying.
    \item Spot-variance correlation: $\rho=-0.7$. 
\end{itemize}
We consider the case of a market maker dealing with $20$ European call options written on that stock where the strike$\times$maturity couples are the elements $(K^j,T^j),j\in \{1,...,20\}$ of the set $\mathcal K\times \mathcal T$, where
\begin{align*}
    \mathcal{K}=\{97,98,99,100\}, \quad  \mathcal{T}=\{0.3 \text{ year} ,0.4 \text{ year}, 0.5 \text{ year}, 0.6 \text{ year}, 0.7 \text{ year}\}. 
\end{align*}
These market parameters provide the implied volatility surface as in Figure \ref{impvolsurf}. 
\vspace{-4mm}
\vspace{-5mm}
We consider mainly in-the-money options with maturity ranging from $3$ to $6$ months so that, due to the influence of both Vanna and Vomma, the Vega of the portfolio changes noticeably and the prices of options are non negligible. \\

We define the following intensity functions:
\begin{align*}
    \Lambda^{j,a}(S,\nu,\delta)=\Lambda^{j,b}(S,\nu,\delta)=\frac{\lambda^j}{1+\exp\big(\alpha+\frac{\beta}{\mathcal{V}_t^j}\delta\big)},
\end{align*}
for $j\in \{1,\dots,N\}$, where $\lambda^j= \frac{252\times 50}{1+0.7\times |S_0-K^j|}\text{ year}^{-1}$, $\alpha=-0.7$, and $\beta=10 \text{ year}^{\frac{1}{2}}$. The choice of $\lambda^j$ corresponds to $50$ requests per day for at-the-money options, and decreases to 13.2 for the most in-the-money options. The choice of $\alpha$ corresponds to a probability of $\frac{1}{1+e^{-0.7}}\approx 66\%$ to trade when the answered quote is the mid-price (i.e $\delta=0$). The choice of $\beta$ corresponds to a probability of $\frac{1}{1+e^{-0.8}}\approx 68\%$ to trade when the answered quote corresponds to an implied volatility $1\%$ better for the client and a probability of $\frac{1}{1+e^{-0.6}}\approx 64\%$ to trade when the answered quote corresponds to an implied volatility $1\%$ worse for the client. \\

We assume transactions of constant size with $z^j = \frac{5\times 10^5}{\mathcal{O}_0^j}$ contracts for option $j$, in other words, the measures $\mu^{j,b},\mu^{j,a}$ are Dirac masses at~$z^j$. This corresponds approximately to $500000$\euro{} per transaction. \\

We finally set $T=0.004$ year (i.e $1$ day), and a risk aversion parameter $\gamma= 2\dot 10^{-5}$\euro{}$^{-1}$. \\

The HJB equation using the constant Vega assumption of \cite{baldacci2019algorithmic} is
\begin{align*}
    0 =  \,\, &\partial_t u(t,\nu,\mathcal{V}^\pi) + a_{\mathbb{P}}(t,\nu)\partial_{\nu}u(t,\nu,\mathcal{V}^\pi) + \frac{1}{2}\nu \xi^2 \partial_{\nu \nu} u(t,\nu,\mathcal{V}^\pi) + \mathcal{V}^\pi \frac{a_{\mathbb{P}}(t,\nu)-a_{\mathbb{Q}}(t,\nu)}{2\sqrt{\nu}}-\frac{\gamma\xi^2}{8} (\mathcal{V}^\pi)^2 \\
    & + \!\!\!\sum_{j\in \{1,\dots,N\}} \!\!\!z^j H^{j,b}\Big(\frac{u(t,\nu,\mathcal{V}^\pi)-u(t,\nu,\mathcal{V}^\pi+z^j \mathcal{V}^j)}{z^j} \Big)  +\!\!\! \sum_{j\in \{1,\dots,N\}} \!\!\! z^j H^{j,a}\Big(\frac{u(t,\nu,\mathcal{V}^\pi)-u(t,\nu,\mathcal{V}^\pi-z^j \mathcal{V}^j)}{z^j} \Big),
\end{align*}
with terminal condition $u(T,\nu,\mathcal{V}^\pi)=0$, and
\begin{align*}
   & \mathcal{V}_t^{\pi} = \sum_{j\in \{1,\dots,N\}} z^j \mathcal{V}^j q_t^{j}, \\
   & H^{j,a/b}(p)=\sup_{\delta^{j,a/b}} \Lambda^{j,a/b}(\delta^{j,a/b})(\delta^{j,a/b}-p). 
\end{align*}

In the case where Vega are not constant, we use the following ansatz:
\begin{align*}
    u(t,S,\nu,q)= \theta^0(t,S,\nu) + q^{\mathbf{T}}\theta^1(t,S,\nu) + q^{\mathbf{T}}\theta^2(t,S,\nu)q,
\end{align*}
where $\theta^0 \in \mathbb{R},\theta^1\in \mathbb{R}^N,\theta^2\in \mathcal{M}_N(\mathbb{R})$. Define
\begin{align*}
    \tilde{\mathcal{L}}(t,S,\nu,\theta) = a_{\mathbb{P}}(t,\nu)\partial_{\nu}\theta(t,S,\nu)+ \frac{1}{2}\nu \xi^2 \partial_{\nu \nu}\theta(t,S,\nu) + \frac{1}{2}\nu S^2 \partial_{SS}\theta(t,S,\nu) + \rho \nu S \xi \partial_{\nu S}\theta(t,S,\nu),
\end{align*}
and assume symmetry of intensity functions, that is $H^{j,b}=H^{j,a}=H^j$, we obtain the following system of coupled PDEs:
\begin{align}\label{sys_pde_matrix}
\begin{cases}
 0 = & \partial_t \theta^0(t,S,\nu) + \tilde{\mathcal{L}}(t,S,\nu,\theta^0)  + 2 \sum\limits_{j\in \{1,\dots,N\}}H^j(S,\nu,0) + 2\sum\limits_{j\in \{1,\dots,N\}}z^j H^{'j}(S,\nu,0)\theta^2_{j,j}(t,S,\nu) \\
&  {}+ \sum\limits_{j\in \{1,\dots,N\}}H^{'' j  }(S,\nu,0)\big(\theta_j^1(t,S,\nu)\big)^2 \\
0 = & \partial_t \theta^1(t,S,\nu) + \tilde{\mathcal{L}}(t,S,\nu,\theta^1)  + \mathcal{V}_t \frac{a_{\mathbb{P}}(t,\nu)-a_{\mathbb{Q}}(t,\nu)}{2\sqrt{\nu}} + 4 \theta^2(t,S,\nu)\text{diag}\Big(H^{''}(S,\nu,0)\Big) \theta^1(t,S,\nu) \\
0 = & \partial_t \theta^2(t,S,\nu) + \tilde{\mathcal{L}}(t,S,\nu,\theta^2) - \frac{\gamma\xi^2}{8}\text{diag}\big(\mathcal{V}_t\big)\frac{\mathbf{1}\mathbf{1}^T}{N}\text{diag}\big(\mathcal{V}_t\big) \\
 &{} + 4 \theta^2(t,S,\nu) \text{diag}\big(H^{''}(S,\nu,0)\big)\theta^2(t,S,\nu),    
\end{cases}
\end{align}
where 
\begin{align*}
    & \mathcal{V}_t = \big(\partial_{\sqrt{\nu}}O^1(t,S,\nu),\dots, \partial_{\sqrt{\nu}}O^N(t,S,\nu)\big)^{\mathbf{T}},  \qquad \mathbf{1}= (1,\dots,1)^{\mathbf{T}} \in \mathbb{R}^N. 
\end{align*}

\begin{figure}[H]
\begin{center}
\begin{subfigure}{.45\textwidth}
 \includegraphics[width=1\textwidth]{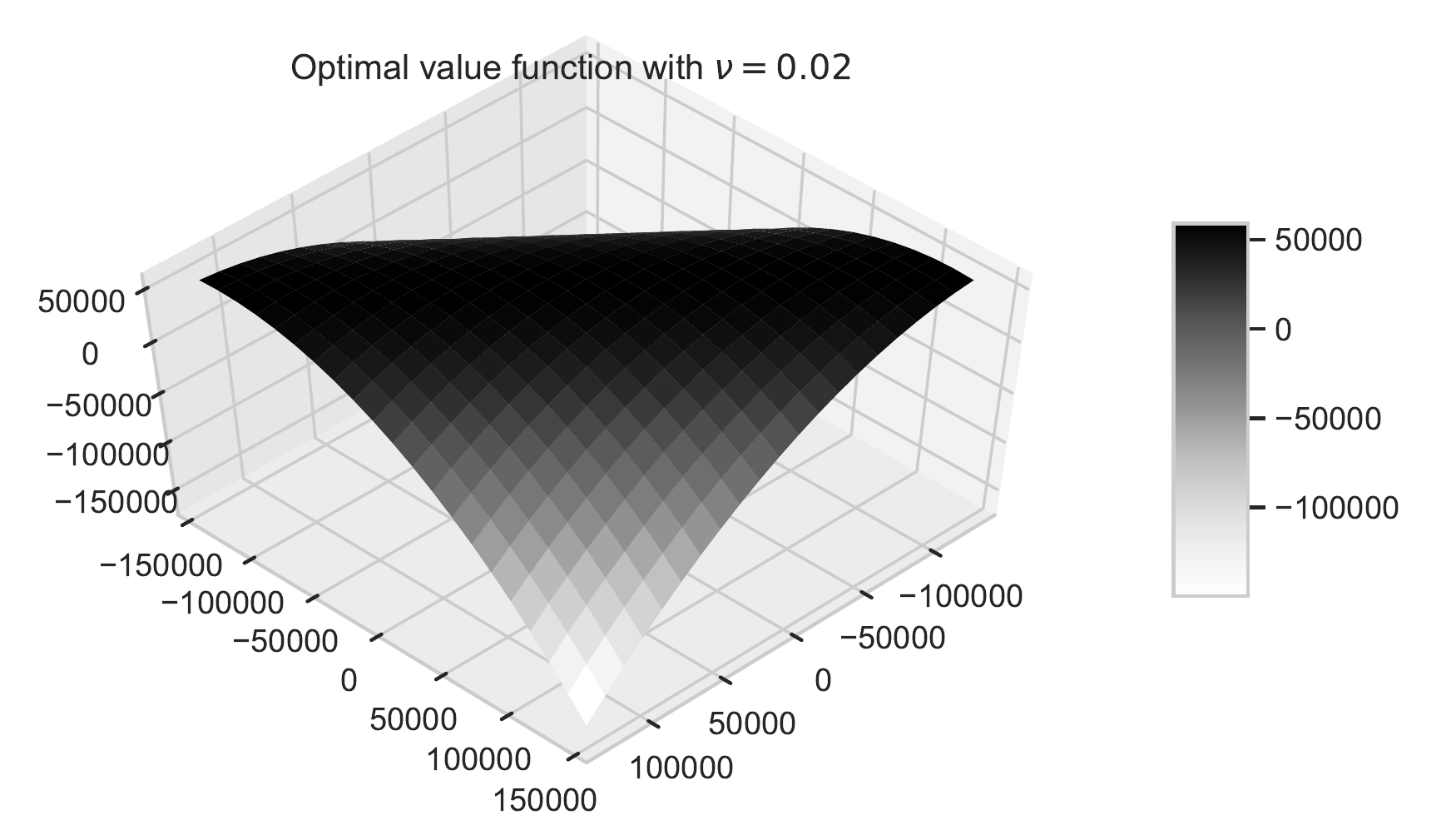}
\end{subfigure}    
\begin{subfigure}{.45\textwidth}
 \includegraphics[width=1\textwidth]{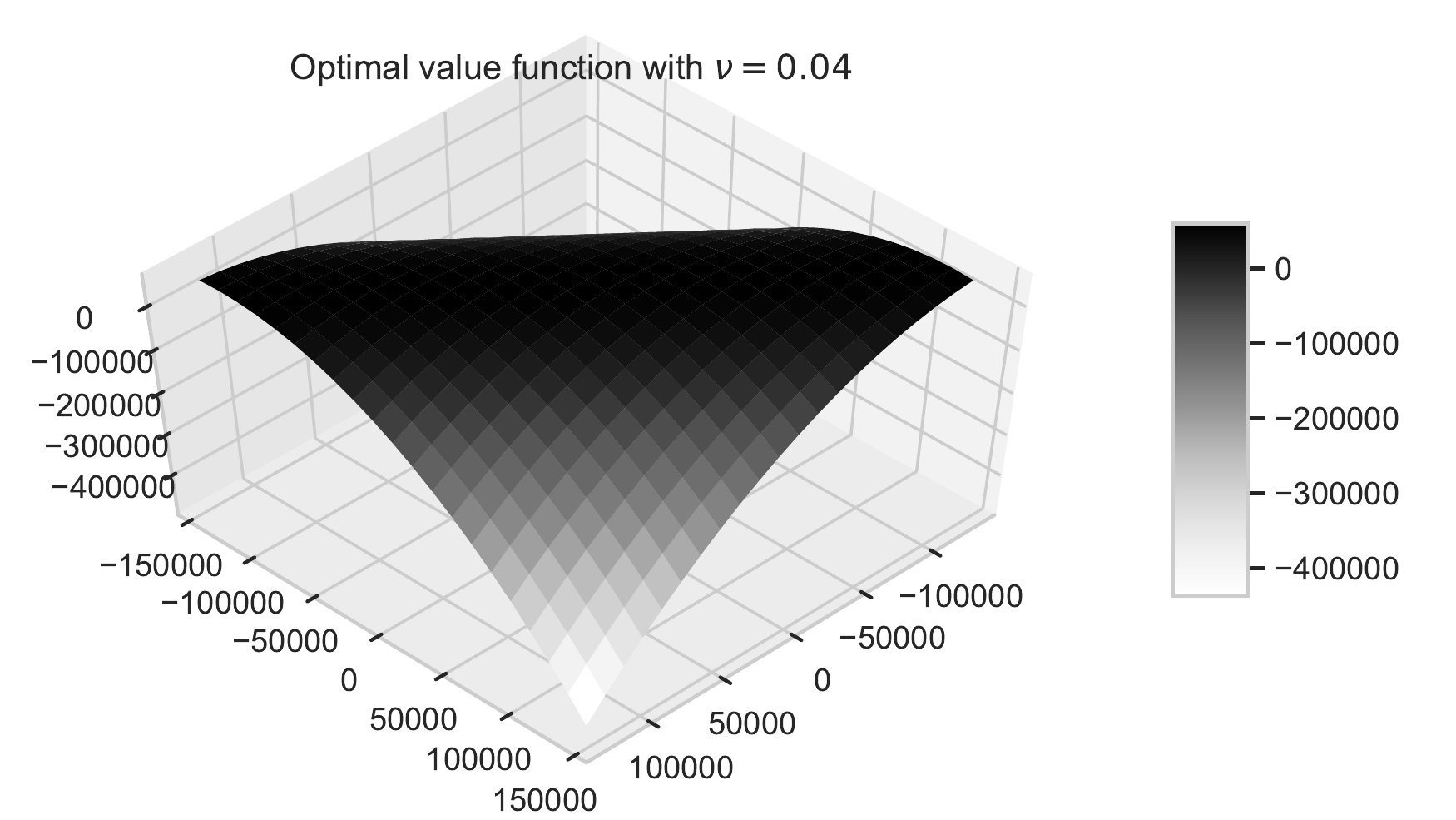}
\end{subfigure}   \\
\begin{subfigure}{.45\textwidth}
\vspace{-5mm}
 \includegraphics[width=1\textwidth]{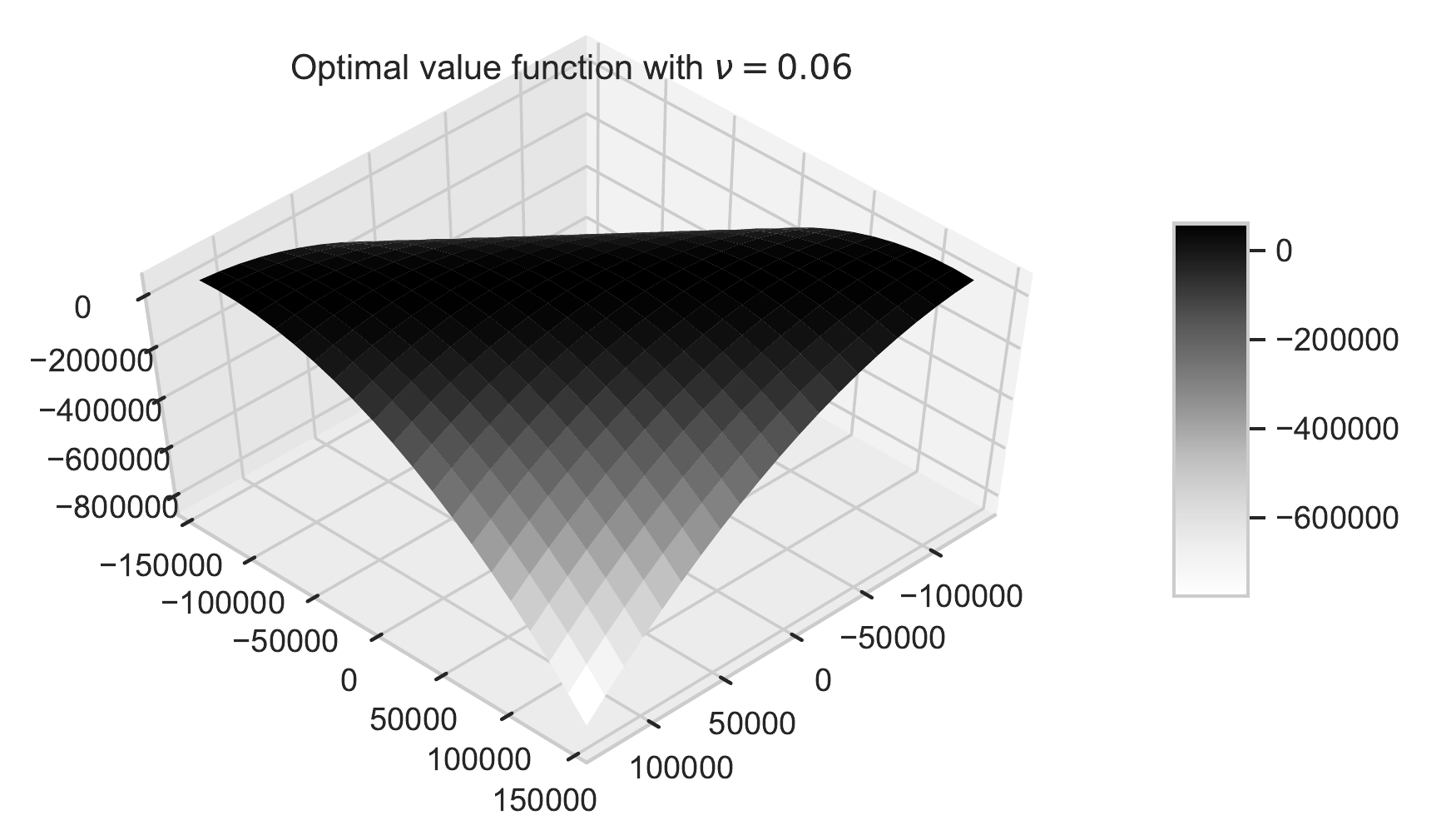}
\end{subfigure}   \\
\vspace{-3mm}
\caption{Value function for different inventories in $(97,0.3)$ and $(98,0.3)$ options, inventories in other options assumed to be equal 0, for different values of $\nu$.}
\label{vf_0_1} 
\end{center}
\end{figure}

\begin{figure}[H]
\begin{center}
\begin{subfigure}{.45\textwidth}
 \includegraphics[width=1\textwidth]{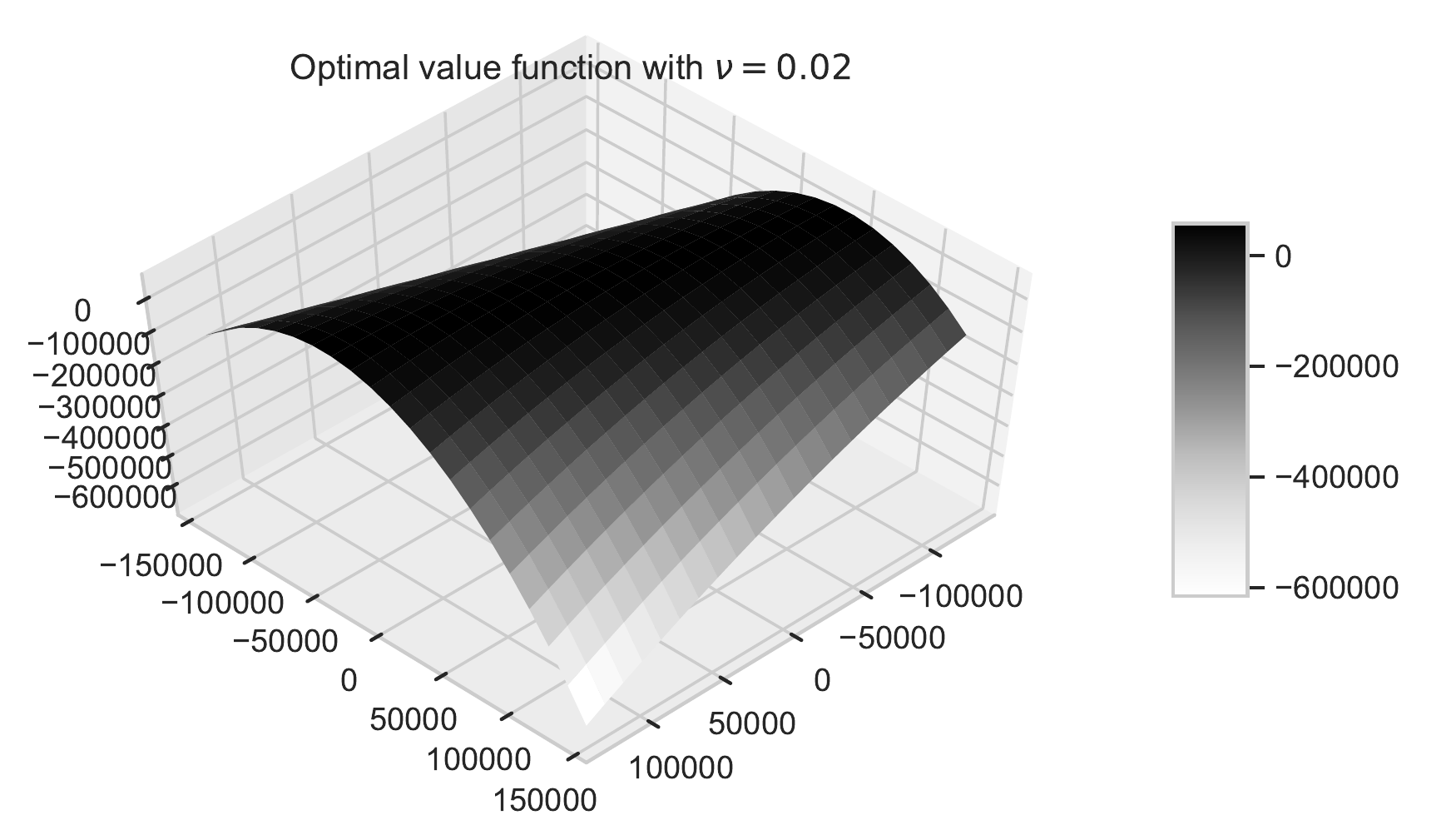}
\end{subfigure}    
\begin{subfigure}{.45\textwidth}
 \includegraphics[width=1\textwidth]{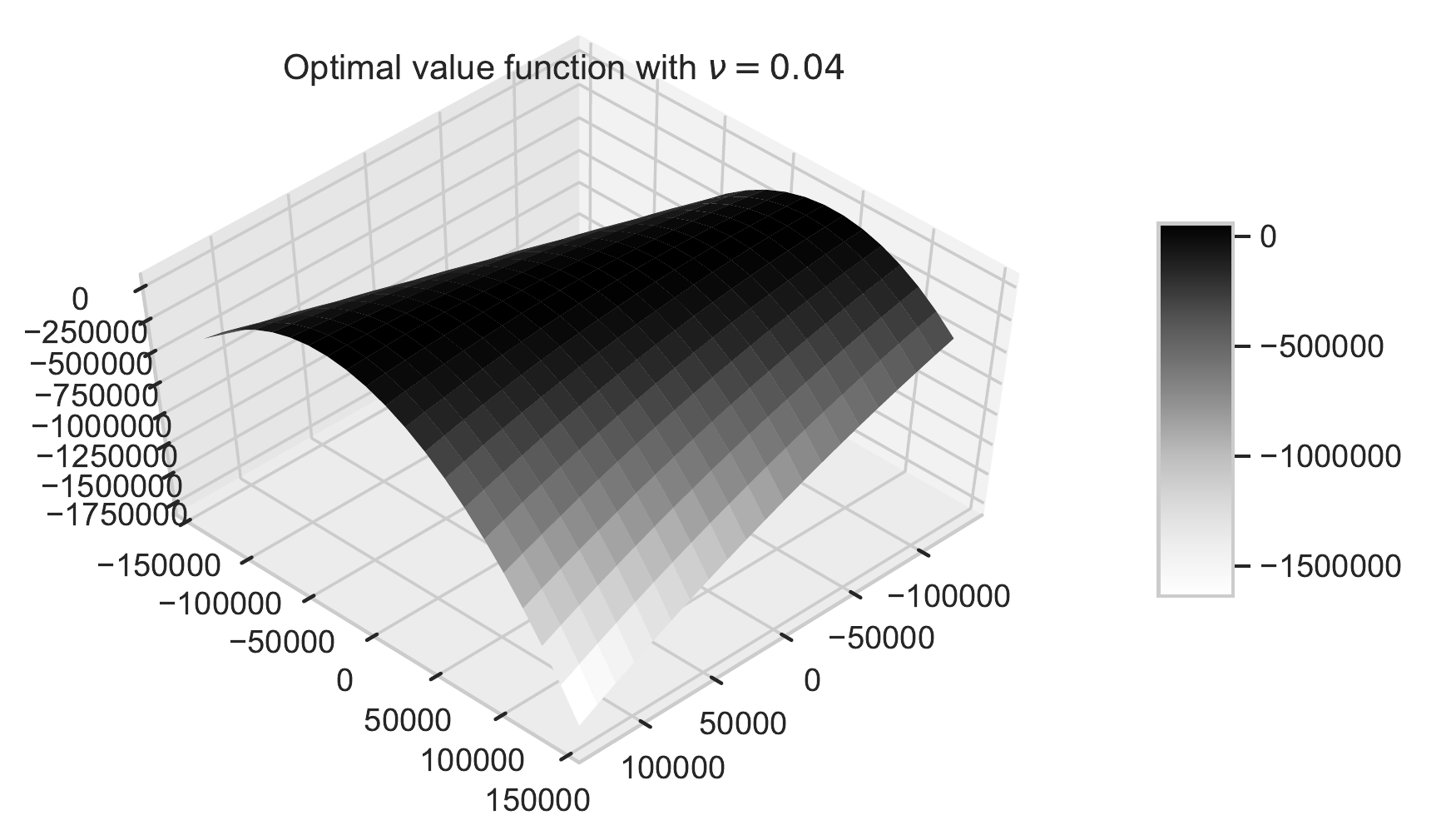}
\end{subfigure}   \\
\begin{subfigure}{.45\textwidth}
\vspace{-5mm}
 \includegraphics[width=1\textwidth]{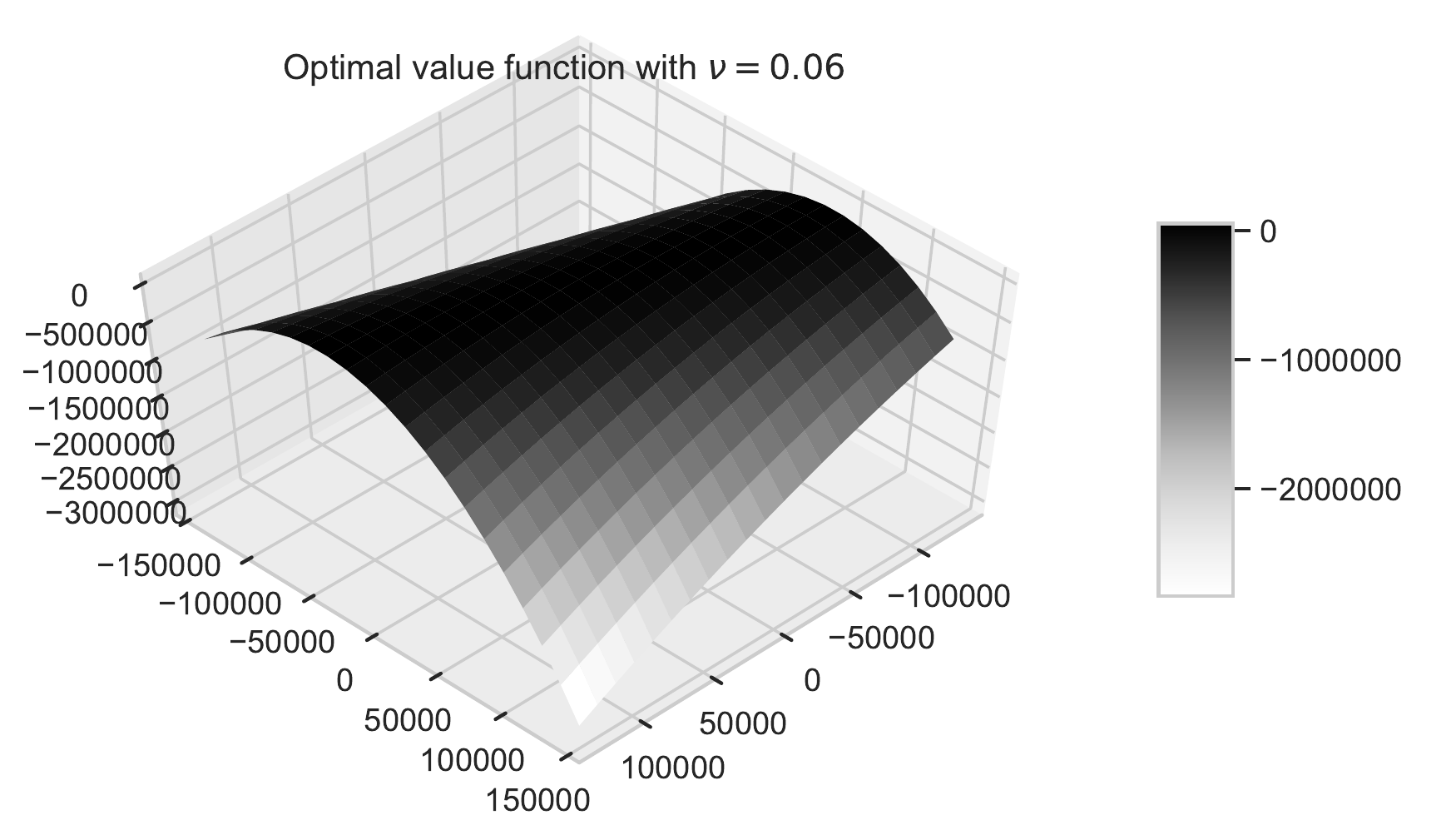}
\end{subfigure} 
\vspace{-3mm}
\caption{Value function for different inventories in $(97,0.3)$ and $(100,0.7)$ options, inventories in other options assumed to be equal 0, for different values of $\nu$.}
\label{vf_0_19} 
\end{center}
\end{figure}

We first show in Figures \ref{vf_0_1} and \ref{vf_0_19} some plots of the value function obtained by solving \eqref{sys_pde_matrix}. 
\vspace{-6mm}

The value function often has higher values on the diagonals. The market maker can compensate a long position in an option with a short position in another one. The values are noticeably lower for higher values of the volatility. \\

\begin{figure}[H]
\begin{center}
\begin{subfigure}{.45\textwidth}
 \includegraphics[width=1\textwidth]{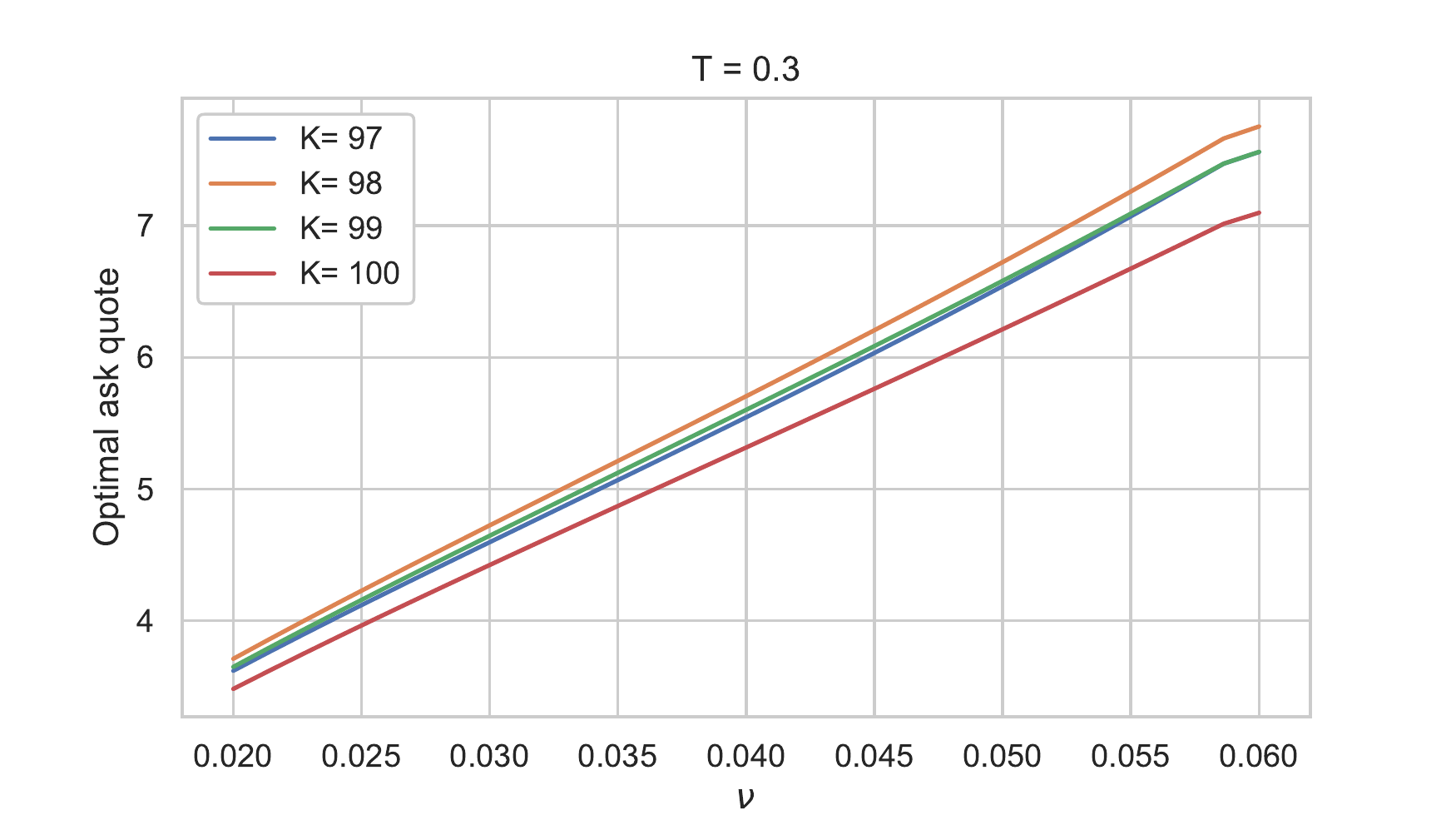}
\end{subfigure}    
\begin{subfigure}{.45\textwidth}
 \includegraphics[width=1\textwidth]{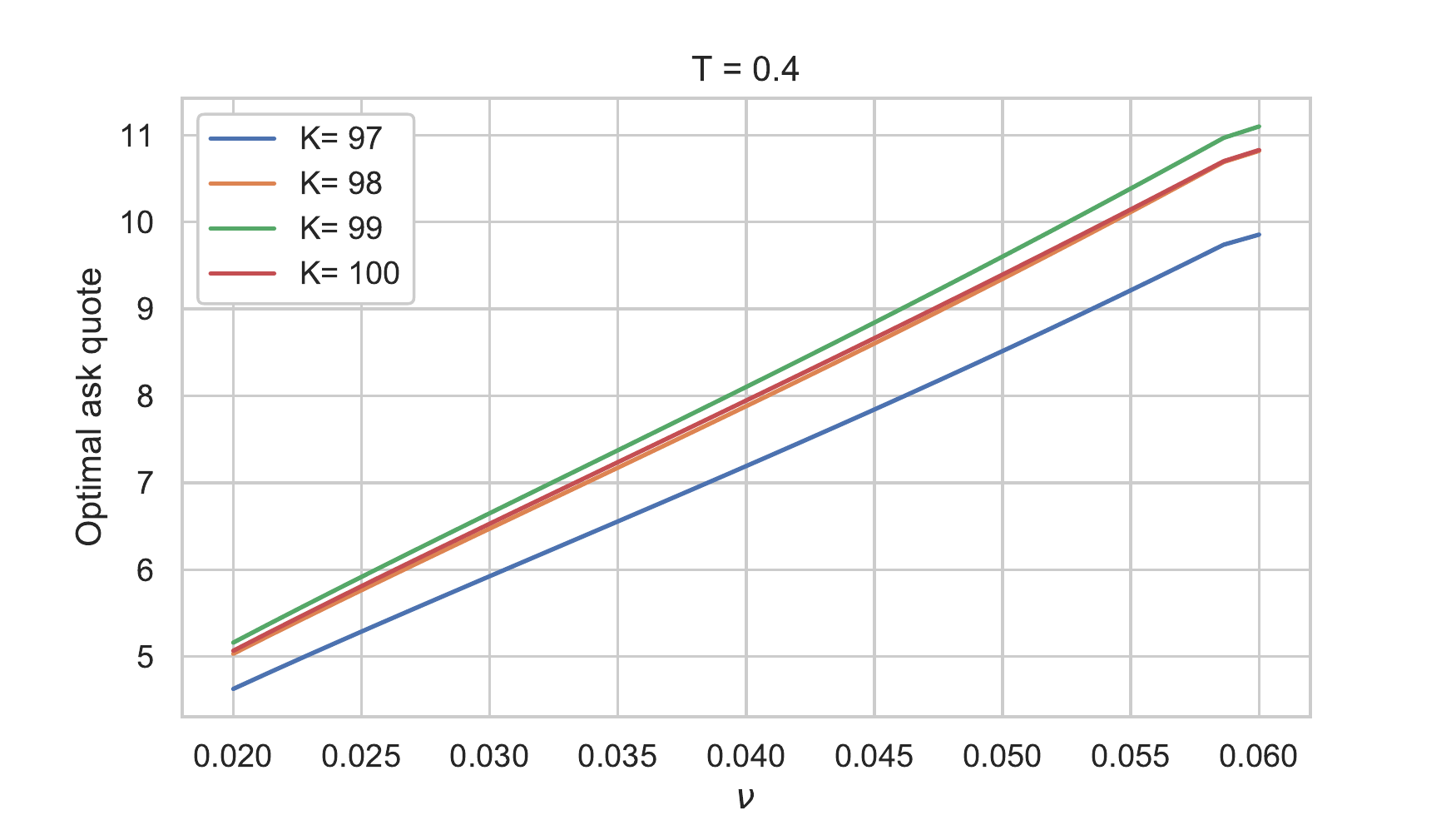}
\end{subfigure}   \\
\begin{subfigure}{.45\textwidth}
 \includegraphics[width=1\textwidth]{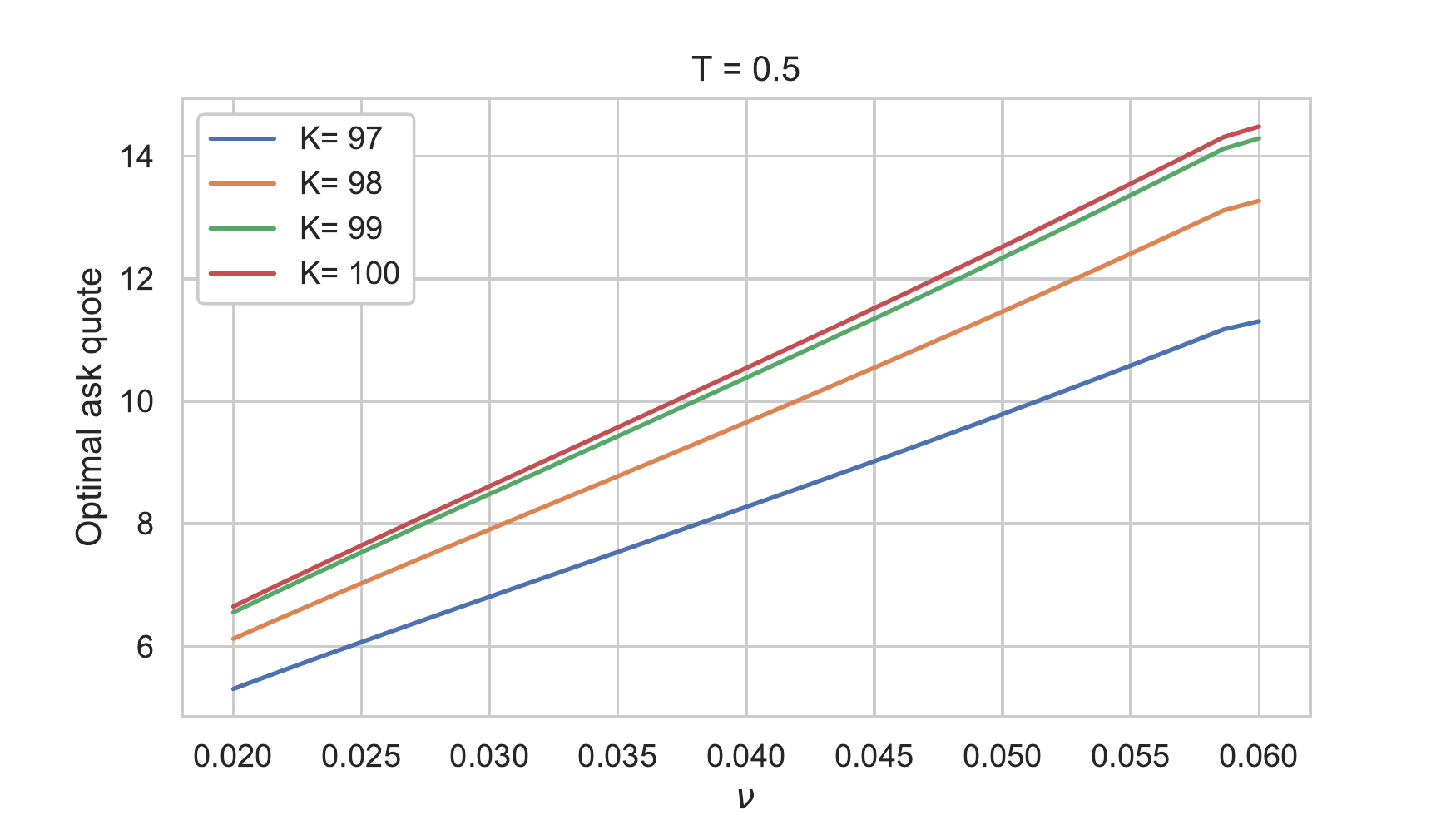}
\end{subfigure} 
\begin{subfigure}{.45\textwidth}
 \includegraphics[width=1\textwidth]{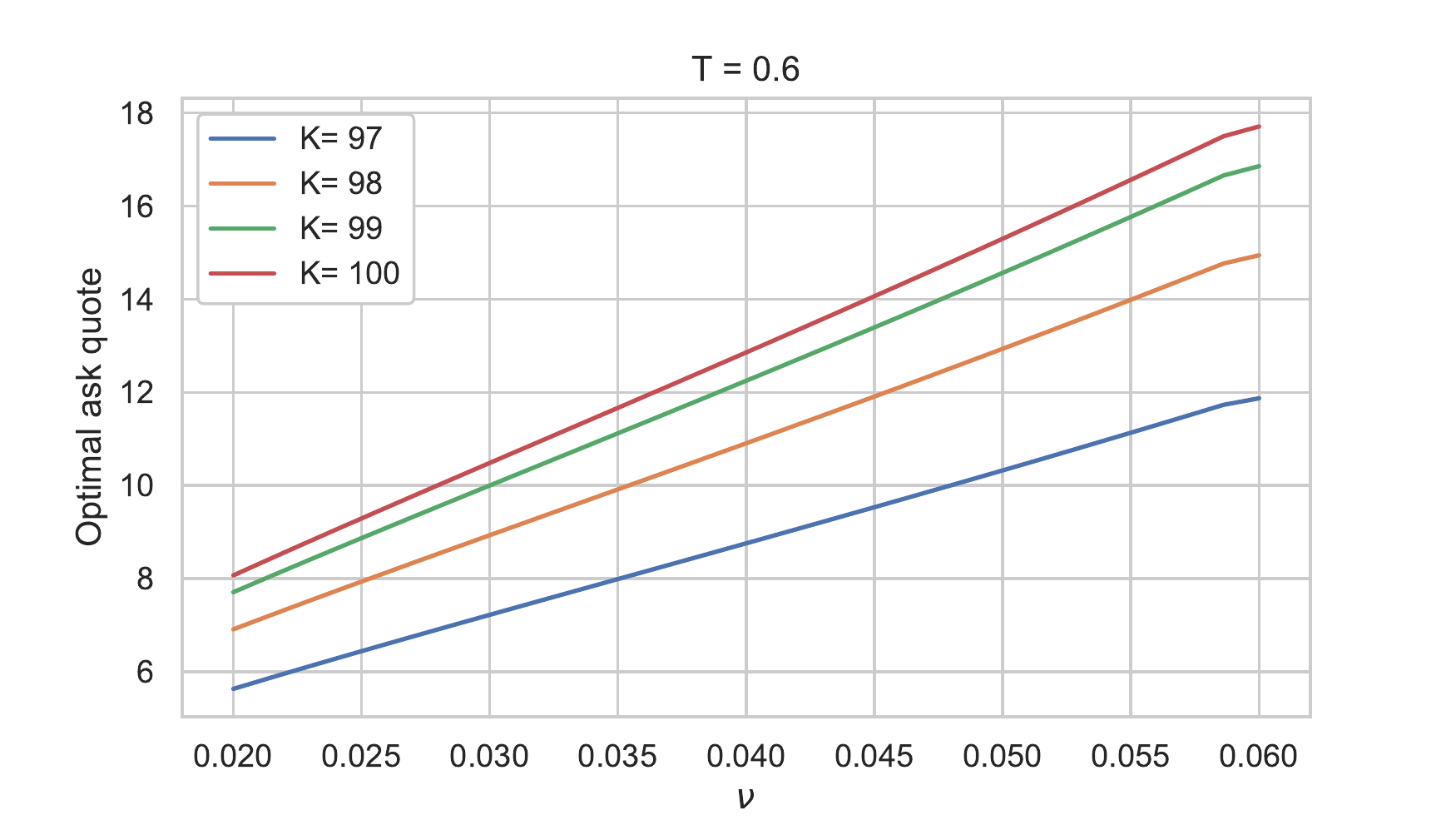}
\end{subfigure} \\
\begin{subfigure}{.45\textwidth}
 \includegraphics[width=1\textwidth]{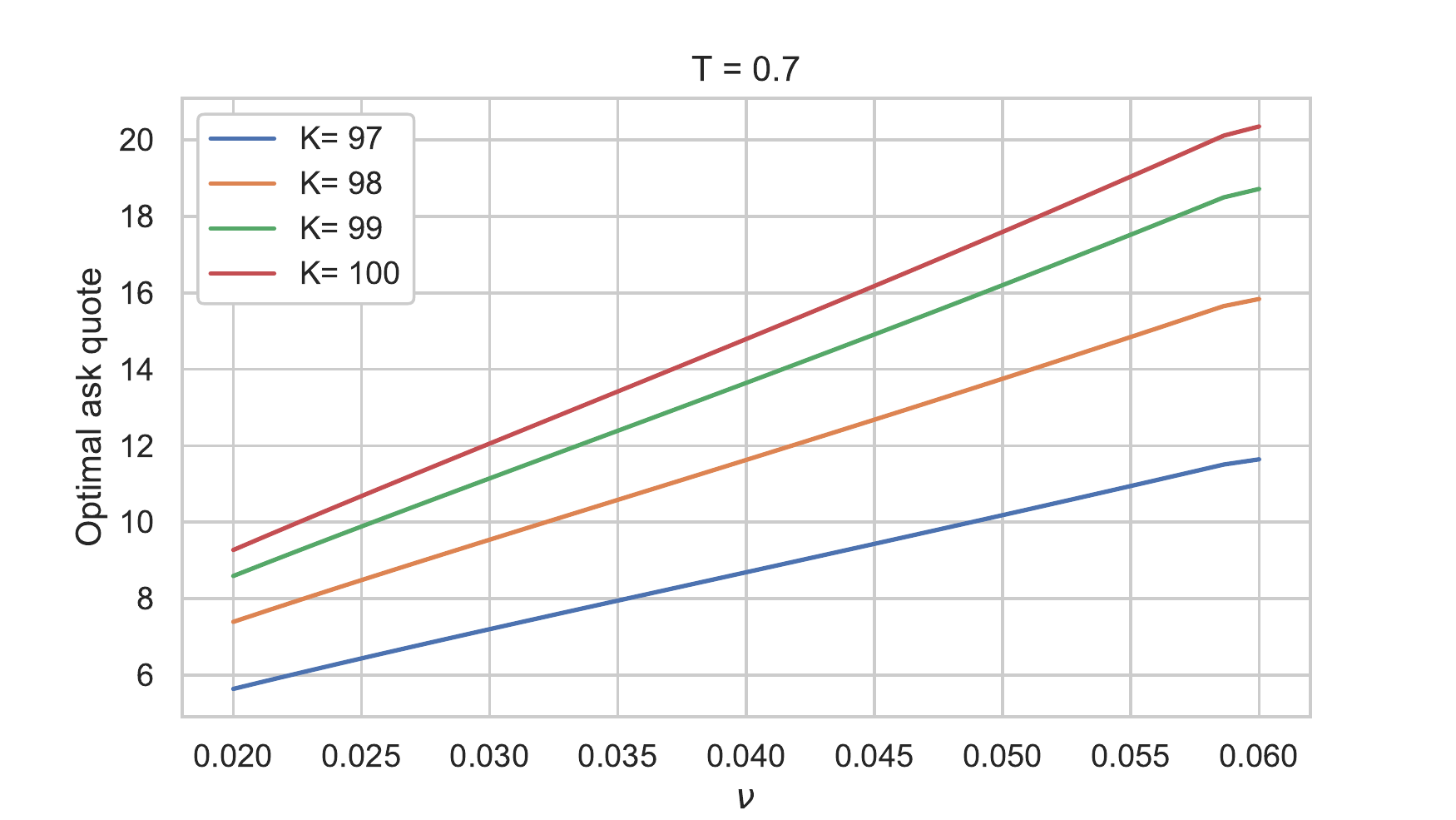}
\end{subfigure} 
\caption{Optimal ask quotes with respect to $\nu$ for different options maturities.}
\label{askdenu} 
\end{center}
\end{figure}

We present in Figure \ref{askdenu} the evolution of the optimal ask quotes with respect to the stochastic volatility for the spot $S=100$.

We observe the usual increasing behavior of the optimal quotes with respect to both maturity and volatility of the underlying. \\

\begin{figure}[H]
\begin{center}
\begin{subfigure}{.45\textwidth}
 \includegraphics[width=1\textwidth]{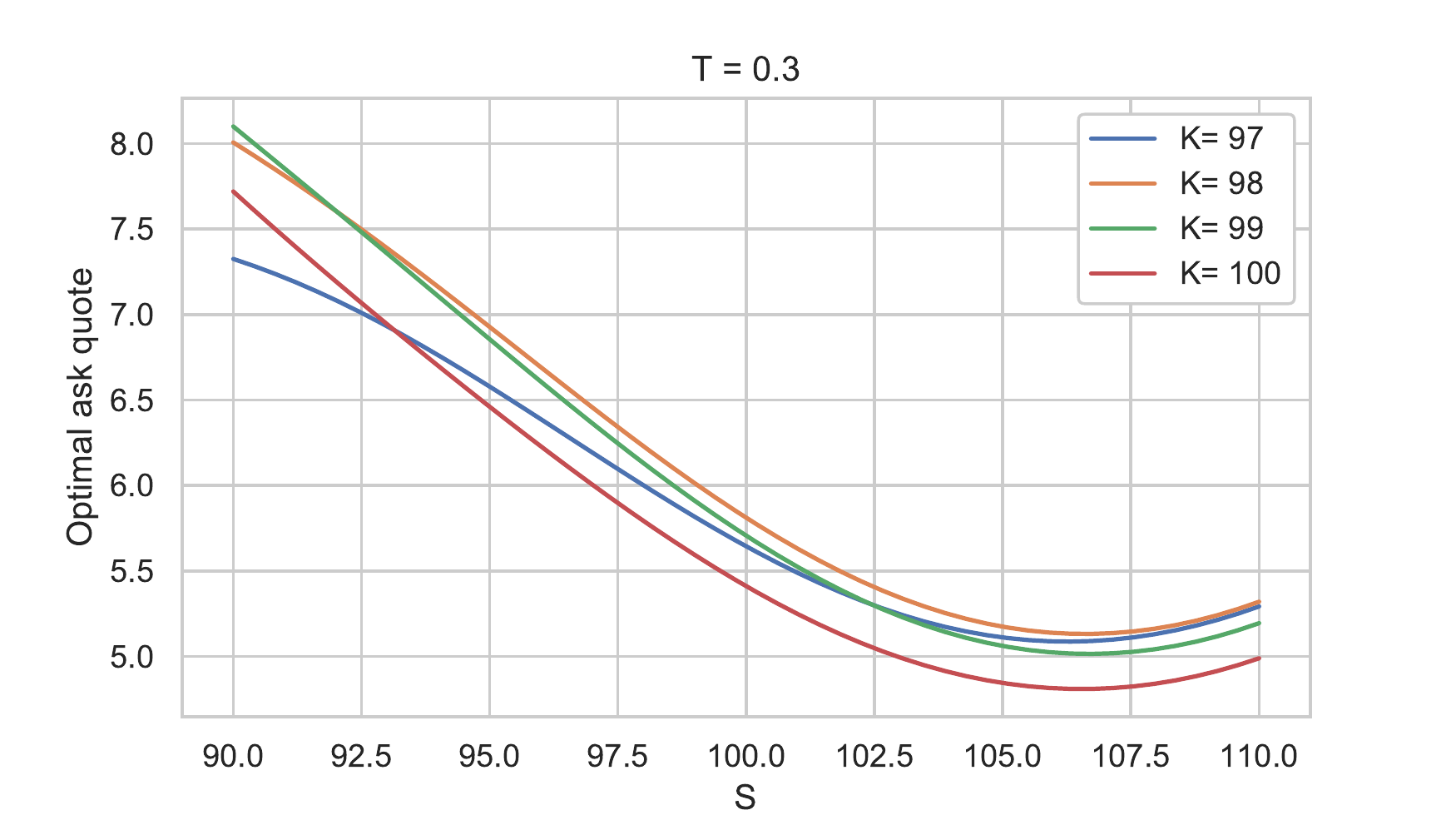}
\end{subfigure}    
\begin{subfigure}{.45\textwidth}
 \includegraphics[width=1\textwidth]{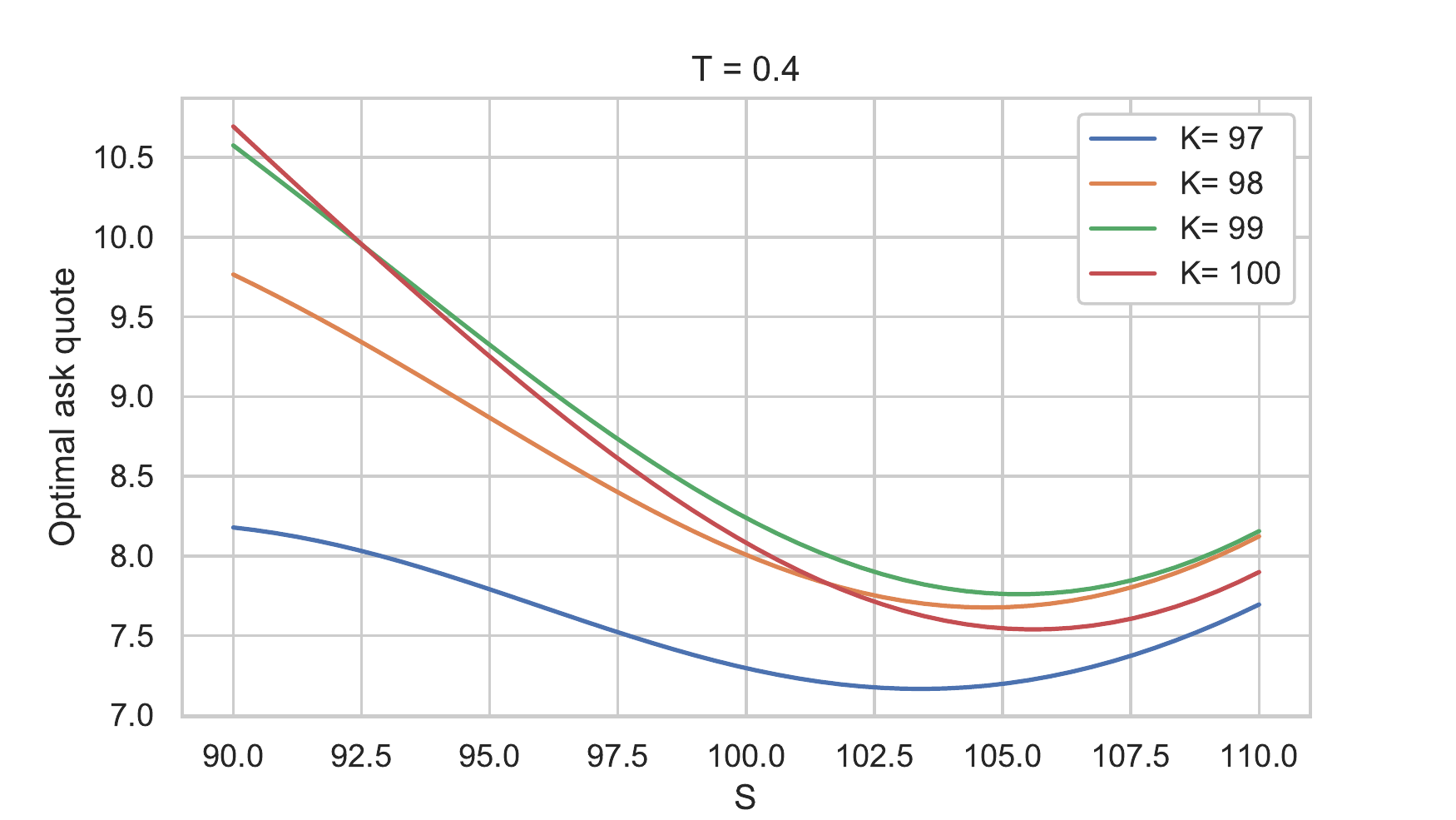}
\end{subfigure}   \\
\begin{subfigure}{.45\textwidth}
 \includegraphics[width=1\textwidth]{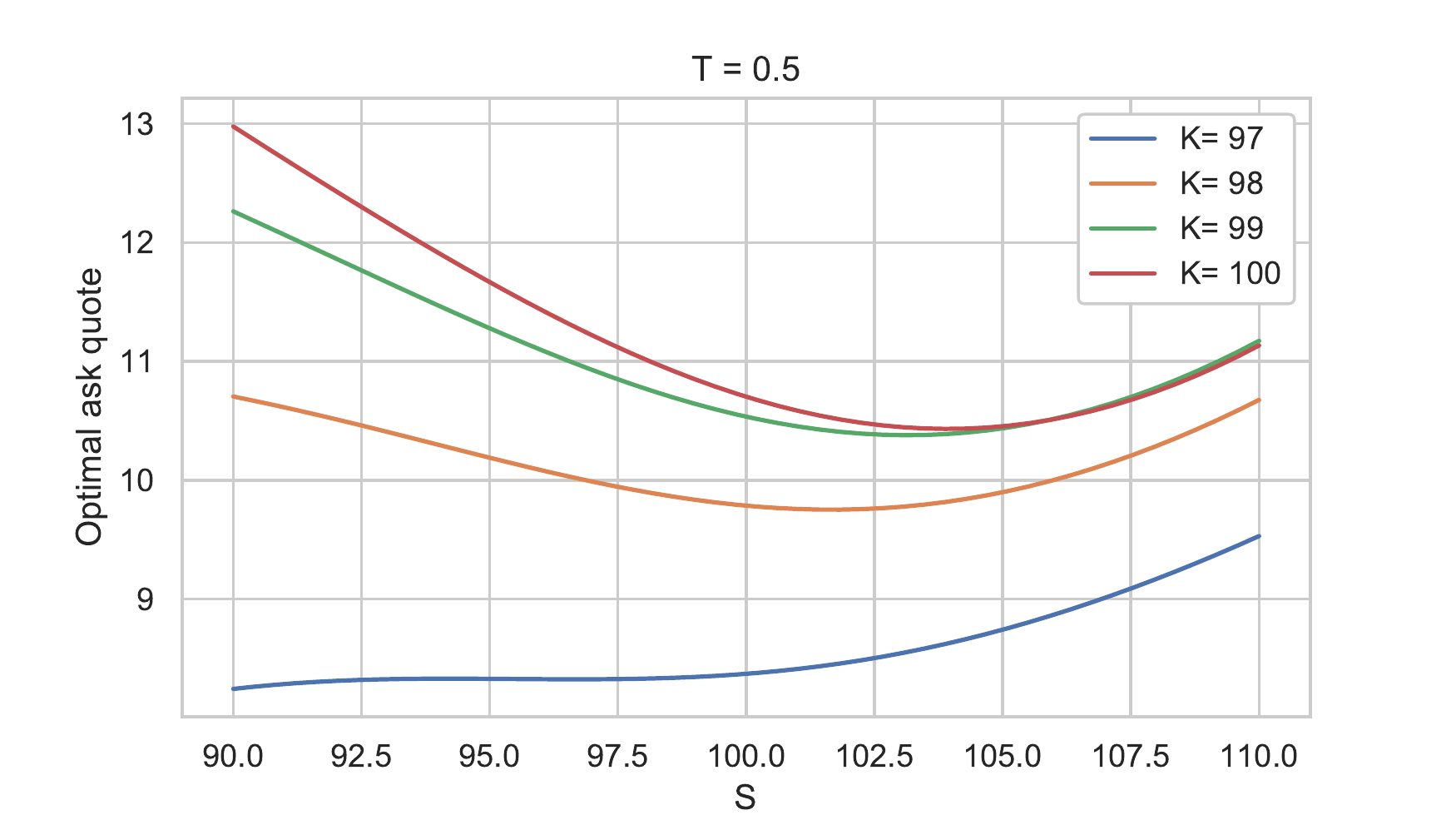}
\end{subfigure} 
\begin{subfigure}{.45\textwidth}
 \includegraphics[width=1\textwidth]{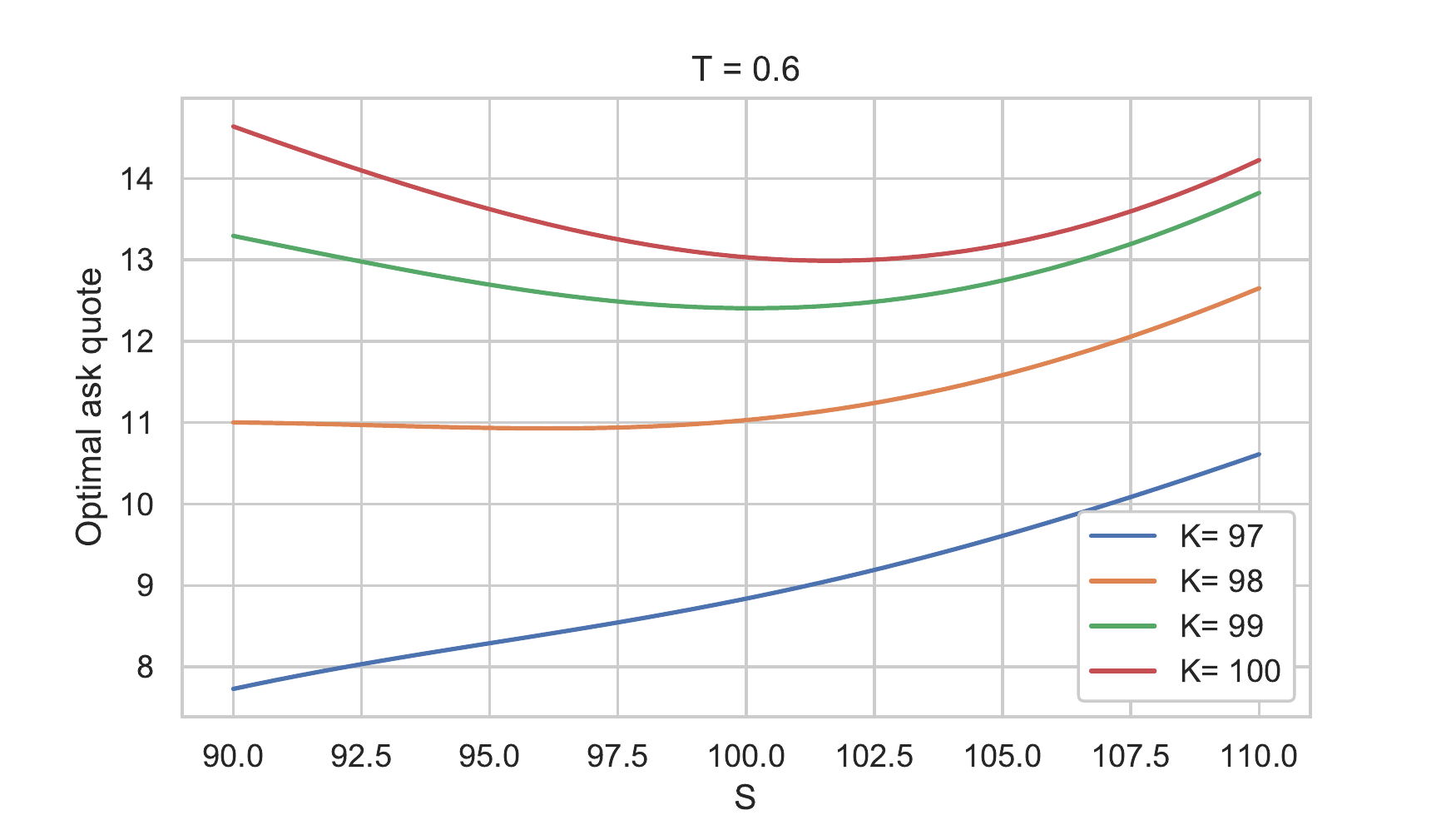}
\end{subfigure} \\
\begin{subfigure}{.45\textwidth}
 \includegraphics[width=1\textwidth]{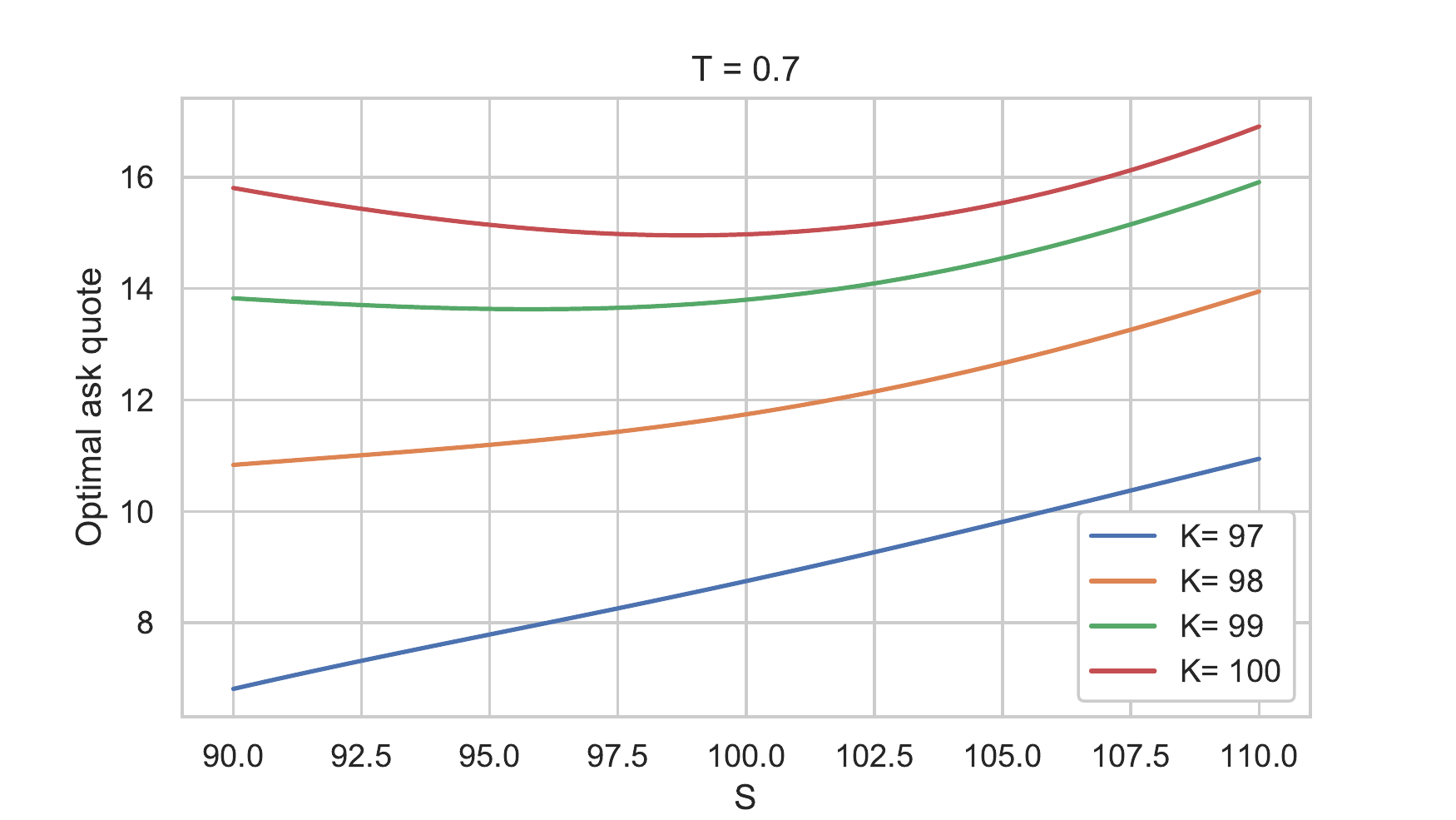}
\end{subfigure} 
\caption{Optimal ask quotes with respect to $S$ for different options maturities.}
\label{askdespot} 
\end{center}
\end{figure}

In Figure \ref{askdespot}, we plot the evolution of the optimal ask quotes with respect to the underlying asset for the volatility $\nu=0.04$. 

\vspace{-5mm}
The behavior of the optimal quotes with respect to the strike depends on the expiry. We can see that the quotes are of the U-shaped nature, the quotes are decreasing in the spot price until some point depending on the strike and the expiry, and then become increasing. The inflection point decreases with the strike decreasing, and conversely for the expiry date. This way we can see that, for example, the quote for the option $(K,T)=(97,0.7)$ is monotonously increasing in the spot price for the considered grid, which is fairly representative of the possible prices during one day. Conversely, for the option $(K,T)=(100,0.3)$ the quote is decreasing for almost all values of the grid.\\  

\begin{figure}[H]
\begin{center}
    \includegraphics[width=0.6\textwidth]{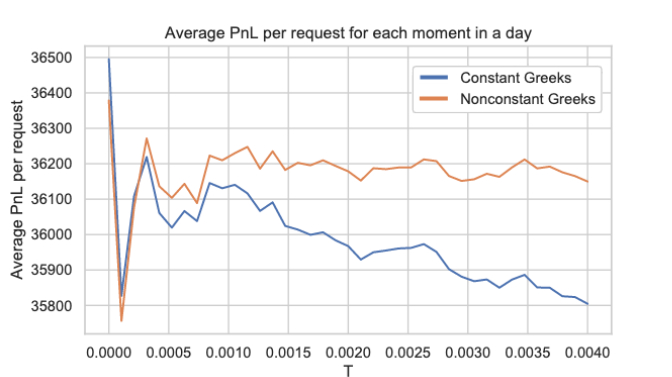}
    \vspace{-3mm}
    \caption{Average PnL per request over the trading day using constant and non-constant Greek approximations.}\label{pnlinaday}
\end{center}
\end{figure}

In Figure \ref{pnlinaday}, we show the average PnL per request of the trader during the day over 1000 simulations, using the constant Greek approximation of \cite{baldacci2019algorithmic} and our algorithm. \\

At the beginning of the trading day, both methods yield a similar PnL per request. Notice that the PnL per request for the method with constant Greek approximation is slightly higher. Indeed the parameters at the beginning of the day correspond to the calibration parameters, and our algorithm is more conservative as it takes into account the risk that the underlying price could change. However, after roughly a tenth of the trading day the method with constant Greek approximations starts to underperform our algorithm. This underperformance increases along the day as the constant Vega approximation becomes less accurate. On the contrary, with our method the PnL per request remains constant: there is no need for recalibration.
\vspace{-3mm}
\vspace{-7mm}

\begin{figure}[H]
\begin{center}
\begin{subfigure}{.45\textwidth}
 \includegraphics[width=1\textwidth]{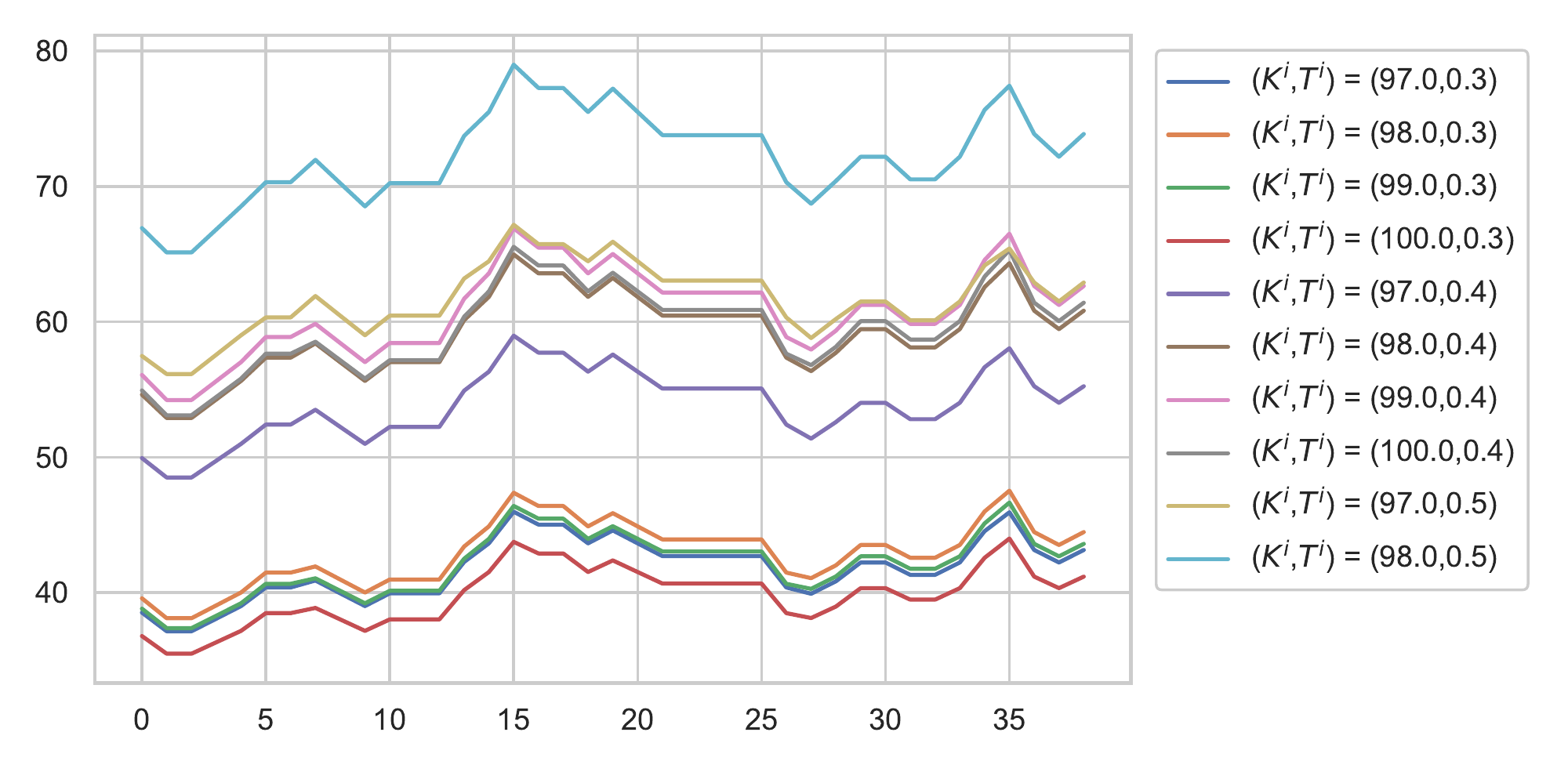}
\end{subfigure}    
\begin{subfigure}{.45\textwidth}
 \includegraphics[width=1\textwidth]{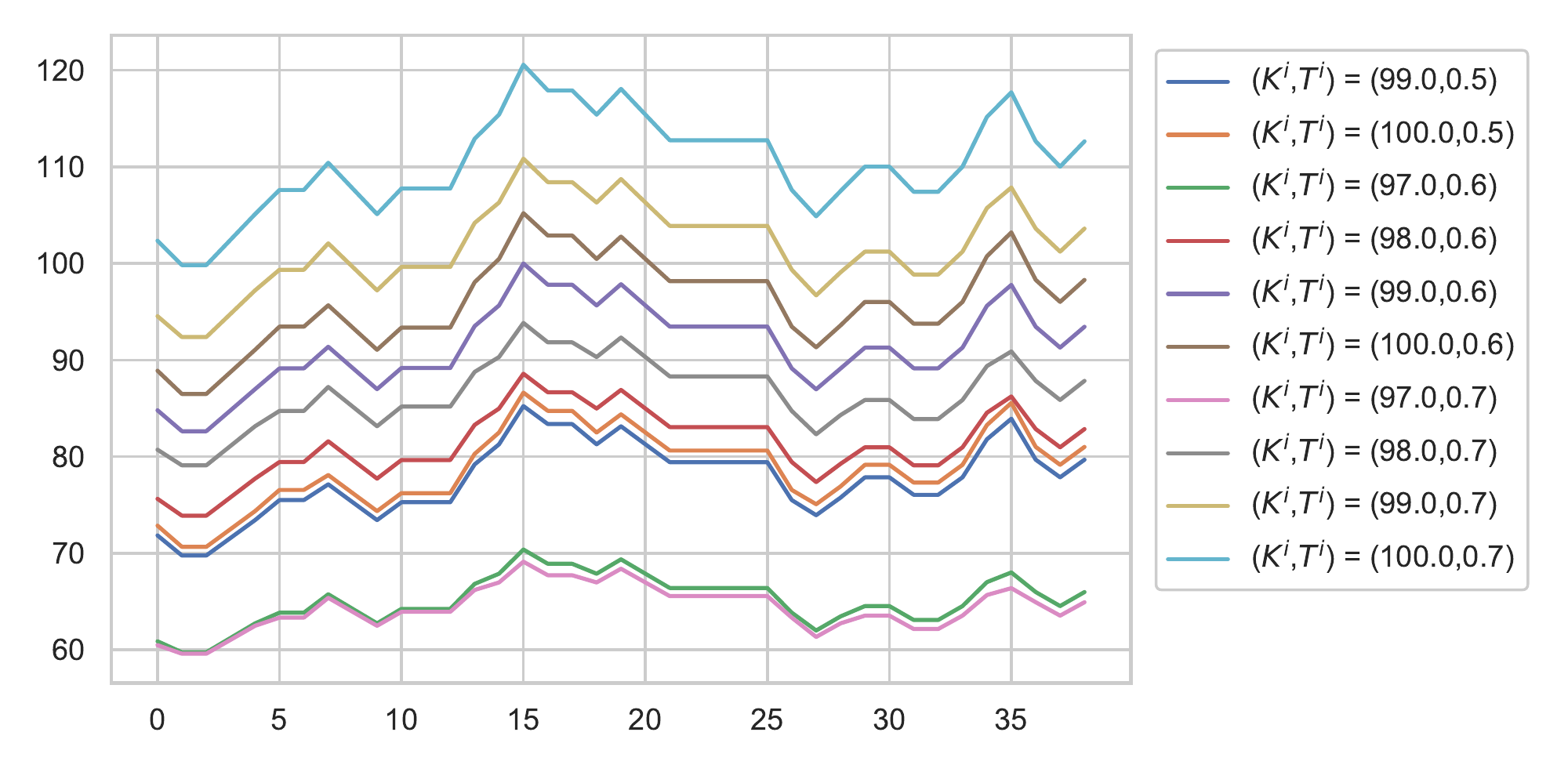}
\end{subfigure}
\vspace{-3mm}
\caption{Example of Vega trajectories.}\label{vegas_example}
\end{center}
\end{figure}

In Figure \ref{vegas_example}, we show one of 1000 simulation examples of the trajectories for the Vega of each option. We see that Vegas for this set of options are changing considerably during the day.
\vspace{-0mm}
\vspace{-7mm}
\begin{figure}[H]
\begin{center}
    \includegraphics[width=0.6\textwidth]{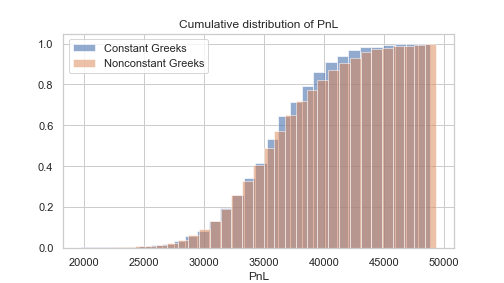}
    \vspace{-2mm}
    \caption{Cumulative distribution functions of the PnL over the trading day using both methods.}\label{cdfpnls}
\end{center}
\end{figure}

Finally, we present in Figure \ref{cdfpnls} the cumulative distribution function of the PnL of the trader using the constant Greek approximation of \cite{baldacci2019algorithmic} and our algorithm. We observe that the tail distribution of the PnL using our non-constant Greek approximation is higher compared to the method in \cite{baldacci2019algorithmic}.

\begingroup
\setcounter{section}{0}
\renewcommand\thesection{Appendix \Alph{section}}   
\section{The market maker's problem for large number of underlyings}

In this appendix, we present the system of low-dimensional PDEs analogous to \eqref{system_pde} for more complex cases such as the market making problem on several underlyings or the case where a number of different options' parameters is large (over one hundred). \\

We can rewrite the system of $(\sum_{i\in\{1,...,d\}}N^i)^2$ equations \eqref{system_pde} on $\theta^2$ as a set of $d^2$ equations by adding the strike and the maturity to the state variables. The same can be applied for the $\theta^1$ equation. This way we obtain a smaller set of equations, though having more dimensions and some non-local terms.\\

Let $\O^i = \big\{(T^{i,j},K^{i,j}), j\in\{1,...N^i\}\big\}$ be the set of parameters of options on the underlying $i\in\{1,...,d\}$ and let us define $\hat\theta^1_i:[0,T]\times\R\times\R_+\times\O^i \to \mathbb{R}$ such that, for all $j\in\{1,...N^i\}$,
\begin{align*}
    \hat\theta^1_i(t,S,\nu,(T^{i,j},K^{i,j})) = \theta^1_{\sum_{l=1}^{i-1} N^l+j}(t,S,\nu).
\end{align*}
Similarly for $i_1,i_2\in\{1,...,d\}$, define $\hat\theta^2_{i_1,i_2}:[0,T]\times\R\times\R_+\times\O^{i_1}\times\O^{i_2}\to \mathbb{R}$ such that, for any $j\in\{1,...N^{i_1}\}$ and $l\in\{1,...N^{i_2}\}$,
\begin{align*}
    \hat\theta^2_{i_1,i_2}(t,S,\nu,(T^{i_1,j},K^{i_1,j}), (T^{i_2,l},K^{i_2,l})) = \theta^2_{\sum_{l=1}^{i_1-1} N^l+j,\sum_{l=1}^{i_2-1} N^l+l}(t,S,\nu).
\end{align*}
Then the system of non-linear PDEs \eqref{system_pde} can be rewritten as 
\begin{align*}
\begin{cases}
0 =& \partial_t \theta^0(t,S,\nu) + \overline{\mathcal{L}}(t,S,\nu,\theta^0)  + 2 \underset{i\in\{1,\dots,d\}}{\sum}\underset{(T,K)\in\O^i}{\sum}H^i(S,\nu,0)(T,K) \\
&{}+ 2\underset{i\in\{1,\dots,d\}}{\sum}\underset{(T,K)\in\O^i}{\sum}\int_{\mathbb R_+}z H^{'i}(S,\nu,0)(T,K)\hat\theta^2_{i,i}\big(t,S,\nu,(T,K),(T,K)\big)\mu^{i,(T,K)}(dz) \\
& + \underset{i\in\{1,\dots,d\}}{\sum}\underset{(T,K)\in\O^i}{\sum}H^{''i}(S,\nu,0)(T,K)\Big(\hat\theta^1_i\big(t,S,\nu,(T,K)\big)\Big)^2 \\
0 =& \partial_t \hat\theta^1_i(t,S,\nu,(\mathcal{T}^1,\mathcal{K}^1)) + \overline{\mathcal{L}}^1\big(t,S,\nu,\hat\theta^1_i,(\mathcal{T}^1,\mathcal{K}^1)\big)  +\mathcal{G}_i(t,S,\nu,(\mathcal{T}^1,\mathcal{K}^1))\\
&{}+ 4 \underset{i_2\in\{1,\dots,d\}}{\sum}\underset{(T,K)\in\O^{i_2}}{\sum}\hat\theta^2_{i,i_2}(t,S,\nu,(\mathcal{T}^1,\mathcal{K}^1),(T,K))H^{''i_2}(S,\nu,0)(T,K) \hat\theta^1_{i_2}\big(t,S,\nu,(T,K)\big) \\
0  =& \partial_t \hat\theta^2_{i_1,i_2}\big(t,S,\nu,(\mathcal{T}^1,\mathcal{K}^1),(\mathcal{T}^2,\mathcal{K}^2)\big) + \overline{\mathcal{L}}^2\big(t,S,\nu,\hat\theta^2_{i_1,i_2} ,(\mathcal{T}^1,\mathcal{K}^1),(\mathcal{T}^2,\mathcal{K}^2)\big) \\
&{}-\frac{\gamma}{2} \mathcal{R}_{i_1}\big(t,S,\nu,(\mathcal{T}^1,\mathcal{K}^1)\big) \Sigma^{\nu,i_1,i_2} \mathcal{R}_{i_2}\big(t,S,\nu,(\mathcal{T}^2,\mathcal{K}^2)\big)\\
& {} + 4 \underset{i_3\in\{1,\dots,d\}}{\sum}\underset{(T,K)\in\O^{i_3}}{\sum}\hat\theta^2_{i_1,i_3}\big(t,S,\nu,(\mathcal{T}^1,\mathcal{K}^1),(T,K)\big) \hat H^{''i_3}(S,\nu,0)(T,K)\hat\theta^2_{i_3,i_2}\big(t,S,\nu,(T,K),(\mathcal{T}^2,\mathcal{K}^2)\big), 
\end{cases}
\end{align*}
where $\big((\mathcal{T}^1,\mathcal{K}^1),(\mathcal{T}^2,\mathcal{K}^2)\big)\in \big(\prod_{i\in \{1,\dots,d\}} \mathbb{O}^i\big)^2$ and, for $j\in\{1,...N^{i_1}\}$, $l\in\{1,...N^{i_2}\}$,
\begin{align*}
    &H^{i}(S,\nu,0)(T^{i,j},K^{i,j}) = H^{i,j}(S,\nu,0),\\ &\mathcal{G}_i(t,S,\nu,(T^{i,j},K^{i,j})) = \mathcal{G}(t,S,\nu)_{\sum_{l=1}^{i-1} N^l+j},\\
    &\mathcal{R}_i(t,S,\nu,(T^{i,j},K^{i,j})) = \mathcal{R}(t,S,\nu)_{\sum_{l=1}^{i-1} N^l+j,i},\\
    &\mu^{i,(T^{i,j},K^{i,j})}=\mu^{i,j},\\
    &\overline{\mathcal{L}}^1\big(t,S,\nu,\hat\theta^1_i,(T^{i,j},K^{i,j})\big) = \overline{\mathcal{L}}(t,S,\nu,\theta^1)_{\sum_{l=1}^{i-1} N^l+j},\\ &\overline{\mathcal{L}}^2\big(t,S,\nu,\hat\theta^2_{i_1,i_2},(T^{i_1,j},K^{i_1,j}),(T^{i_2,l},K^{i_2,l})\big) = \overline{\mathcal{L}}(t,S,\nu,\theta^2)_{\sum_{l=1}^{i_1-1} N^l+j,\sum_{l=1}^{i_2-1} N^l+l}.
\end{align*} 
In particular, if $d=1$, $\hat\theta^1$ and $\hat\theta^2$ are solutions of non-local PDEs in dimensions 5 and 7 respectively. The observed regularity of the solution with respect to the strike and expiry implies that the high-dimensional PDEs can be solved, for example, by a non-local variant of the Deep Galerkin Method, see~\cite{hirsa2020unsupervised, Sirignano_2018}.

\bibliographystyle{abbrv}
\bibliography{biblio.bib}

\end{document}